\documentclass[trackchanges]{aastex62}
\shorttitle{NIR spectra of SN~2014J in the nebular phase}
\shortauthors{Diamond et al.}
\usepackage{graphicx}
\usepackage[version=3]{mhchem}
\usepackage{aas_macros}
\usepackage{natbib}
\newcommand{\unit}[1]{\ensuremath{\, \mathrm{#1}}}
\newcommand{\mch}{M_{\rm Ch}}
\newcommand{\co}{\ce{C}-\ce{O}}
\newcommand{\myF}{\mathscr{F}}
\usepackage{mathrsfs}
\usepackage[inline]{enumitem}
\received{March 5, 2018}
\revised{May 8, 2018}
\accepted{May 9, 2018}

\turnoffedit

\begin{document}

\title{Near-infrared spectral evolution of the Type Ia supernova 2014J in the nebular phase: implications for the progenitor system}

\author[0000-0002-0805-1908]{T.~R. {Diamond}}
\affiliation{NASA Postdoctoral Program Fellow}
\affiliation{Laboratory of Observational Cosmology, Code 665, NASA Goddard Space Flight Center, Greenbelt, MD 20771, USA}

\author[0000-0002-4338-6586]{P. {Hoeflich}}
\affiliation{Department of Physics, Florida State University, Tallahassee, FL 32306, USA}

\author[0000-0003-1039-2928]{E.~Y. {Hsiao}}
\affiliation{Department of Physics, Florida State University, Tallahassee, FL 32306, USA}

\author[0000-0003-4102-380X]{D.~J. {Sand}}
\affiliation{Department of Astronomy / Steward Observatory, 933 North Cherry Avenue, Rm. N204, Tucson, AZ 85721-0065, USA}

\author[0000-0003-1440-9897]{G. {Sonneborn}}
\affiliation{Laboratory of Observational Cosmology, Code 665, NASA Goddard Space Flight Center, Greenbelt, MD 20771, USA}

\author[0000-0003-2734-0796]{M.~M. {Phillips}}
\affiliation{Las Campanas Observatory, Carnegie Observatories, Casilla 601, La Serena, Chile}

\author{B. {Hristov}}
\affiliation{Department of Physics, Florida Agricultural \& Mechanical University, Tallahassee, FL 32307, USA}

\author[0000-0001-6661-2243]{D.~C. {Collins}}
\affiliation{Department of Physics, Florida State University, Tallahassee, FL 32306, USA}

\author{C. {Ashall}}
\affiliation{Department of Physics, Florida State University, Tallahassee, FL 32306, USA}

\author{G.~H. {Marion}} 
\affiliation{Department of Astronomy, University of Texas, Austin, TX 78705, USA}

\author[0000-0002-5571-1833]{M. {Stritzinger}}
\affiliation{Department of Physics and Astronomy, Aarhus University, Ny Munkegade 120, DK-8000 Aarhus C, Denmark}

\author[0000-0003-2535-3091]{N. {Morrell}}
\affiliation{Las Campanas Observatory, Carnegie Observatories, Casilla 601, La Serena, Chile}

\author{C.~L. {Gerardy}}
\affiliation{Department of Physics, University of North Carolina, Charlotte, NC 28223, USA}

\author{R.~B. {Penney}}

\correspondingauthor{T.~R. Diamond}
\email{tiaradiamond@gmail.com}

\begin{abstract}
As the closest Type Ia supernova in decades, SN 2014J provides a unique opportunity for detailed investigation into observational signatures of the progenitor system, explosion mechanism, and burning product distribution. 
We present a late-time near-infrared spectral series from Gemini-N at $307-466\unit{days}$ after the explosion.
Following the $H$-band evolution probes the distribution of radioactive iron group elements, the extent of mixing, and presence of magnetic fields in the expanding ejecta.
Comparing the isolated $1.6440\unit{\mu m}$ [\ion{Fe}{2}] emission line with synthetic models shows consistency with a Chandrasekhar-mass white dwarf of $\rho_c=0.7\times10^9\unit{g}\unit{cm^{-3}}$ undergoing a delayed detonation.
The ratio of the flux in the neighboring $1.54\unit{\mu m}$ emission feature to the flux in the $1.6440\unit{\mu m}$ feature shows evidence of some limited mixing of stable and radioactive iron group elements in the central regions.
Additionally, the evolution of the $1.6440\unit{\mu m}$ line shows an intriguing asymmetry.
When measuring line-width of this feature, the data show an increase in line width not seen in the evolution of the synthetic spectra, corresponding to $\approx1{,}000\unit{km}\unit{s^{-1}}$, which could be caused by a localized transition to detonation producing asymmetric ionization in the ejecta. 
Using the difference in width between the different epochs, an asymmetric component in the central regions, corresponding to approximately the inner $2\times10^{-4}$ of white dwarf mass suggests an off-center ignition of the initial explosion and hence of the kinematic center from the chemical center.
Several additional models investigated, including a \ce{He} detonation and a merger, have difficulty reproducing the features seen these spectra.
\end{abstract}



\section{Introduction}
\label{sec:intro}

Type Ia supernovae (SNe Ia) are events that are important for many studies in astronomy, with one of the foremost being high-precision cosmology \citep{phillips93,hamuy95,riess95,hamuy96a,goldhaber98,phillips99,burns14}.
SNe Ia are generally understood to be the thermonuclear explosion of a carbon-oxygen (\co) white dwarf (WD).
However, there is not a consensus on which progenitor channel or channels lead to a normal SN Ia and which lead to subclasses like \edit1{subluminous (eg. 1991bg), over-luminous (eg. 1991T), very over-luminous super-Chandrasekhar (eg. 2009dc), and SNe Iax (eg. 2002cx); a useful summary of observational constraints is presented in \citet{maeda16}.}
The proposed channels include the \textit{single degenerate} (SD) system, in which the \co\, WD has one or more non-degenerate companions, and the \textit{double degenerate} (DD) system, involving a degenerate companion \citep{hoyle60,whelan73,iben84,webbink84,branch95,wang12}.
Starting from these systems, the thermonuclear explosion may be triggered in several different ways, as outlined below.

\textit{Chandrasekhar mass ($\mch$) Explosion:} accretion from the companion onto the WD with the ignition occurring at or near the center of the WD once the mass of the WD is close to $\mch$ (can be either SD or DD system).
This is typically followed by a short deflagration burning phase switching to a detonation \citep{nomoto84}.
The accreting material can be hydrogen or helium from a non-degenerate companion or carbon, oxygen, or helium from a degenerate companion \citep{hoeflich13}.
\textit{Double Detonation:} accretion from the companion onto the WD (typically in a SD system) with the ignition occurring in a thin helium shell of the WD, creating an inward detonation shock that triggers an outward carbon detonation \citep{nomoto82a,nomoto82b}.
\textit{Dynamical and Violent Mergers:} during an in-spiral of two \co\, WDs, a detonation is triggered because of the high temperature from the merger process \citep{pakmor10,pakmor11,pakmor12,shen12,sato15,tanikawa15,liu16}.
\textit{Collision:} a head-on collision of two \co\, WDs in a triple, or larger, system or in a globular cluster, producing the explosion as a direct result of the shock \citep{raskin09,kushnir13,dong15}.

It is not yet clear if these channels contribute to most or all of the normal SNe Ia.
If the main population that contributes to SNe Ia changes with increasing redshift, that will systematically affect cosmological parameter estimates.
With the data coming out of current transient surveys and the much-larger data sets expected from next generation space-based and ground-based telescopes, having tools to discriminate between progenitor channels and explosion scenarios is extremely important.

Deflagration fronts are intrinsically multi-dimensional and strong mixing by Rayleigh-Taylor (RT) instabilities is predicted by hydrodynamical calculations \citep{zeldovich70,khokhlov95,niemeyer95,reinecke99,lisewski00,gamezo03,gamezo05,roepke06,fink14}.
However, detailed studies of well-observed SNe~Ia suggest there is an additional process at work by which mixing is largely suppressed, and a discussion of this can be found in \citet[Appendix C]{gall17}.
Based on these considerations, we will use spherically-symmetric models in this analysis.

In this work, we present an analysis of late-time NIR spectroscopy of SN 2014J using the [\ion{Fe}{2}] emission line at $1.6440\unit{\mu m}$ to probe the initial conditions of the white dwarf.
A short review of the discovery and many followup observations of SN 2014J, including implications for the progenitor system and explosion scenario, will be presented in Section~\ref{sec:14J}.
Details of the observations and an analysis of the NIR spectra in the context of a $\mch$ explosion is presented in Sections~\ref{sec:obs-red} and \ref{sec:analysis}.
A discussion of our results and their implications, followed by conclusions, will be presented in Section~\ref{sec:results}.


\section{SN~2014J and observational evidence for the progenitor system}
\label{sec:14J}

SN~2014J eluded detection by major SN surveys and was instead discovered approximately a week after explosion by a professor with several students \citep{fossey14} and shortly after classified as a SN Ia by \citet{cao14}.
Although SN~2014J was discovered relatively late considering the proximity of its host galaxy, M82, pre-discovery images from surveys do exist and data of the early-time rise of the SN have been analyzed by \citet{zheng14} and \citet{goobar14,goobar15}.
Because of its closeness at under $4\unit{Mpc}$ \citep{dalcanton09}, the post-discovery light curve and spectral evolution of this SN have been well-observed in many different wavelength regions \citep[see][and many others]{ashall14,foley14b,margutti14,pereztorres14,jack15,marion15,siverd15,telesco15,galbany16,sand16,srivastav16}.

Fitting the early rise of the SN light curve, \citet{zheng14} find the best value for the time of first light to be 2014 January $14.75\pm0.21$ UT (JD $2456672.25\pm0.21$), which is consistent with other analyses \citep{ashall14,marion15}, and will be used as the explosion date for SN~2014J in this work.
However, \citet{zheng14} and \citet{goobar15} find that the rise of SN~2014J is not consistent with the $t^2$ model that describes a simple expanding fireball nor with the expected power-law relationship for shock-heated emission from a companion star; rather, a changing power-law provided the best fit of the SN~2014J data.

Spectroscopic observations to look at the abundance distribution in the outer layers of the ejecta by \citet{ashall14} and \citet{marion15} show consistency with models of a $\mch$ explosion with a deflagration to detonation transition.
However, X-ray observations by \citet{margutti14} and radio observations by \citet{pereztorres14} seem to rule out most SD channels for the SN based on mass-loss limits.

Looking at NIR spectra during the transition between the photospheric and nebular phases (about $100\unit{days}$ after the explosion), \citet{sand16} limit the amount of swept-up hydrogen from a companion star to $\approx 0.1\unit{M_{\odot}}$.
\citet{lundqvist15} found no evidence for a non-degenerate companion star in late-phase optical observations of SN~2014J, putting a hydrogen mass limit of $0.0085\unit{M_{\odot}}$ and a helium mass limit of $0.005\unit{M_{\odot}}$ on the amount of material accreted from a companion star.
However, these limits are compatible with either hydrogen or helium companion stars if the orbital separation is large, in addition to a degenerate companion star or the WD merger scenario \citep{lundqvist15}.
In the analysis presented here, we will compare the late-time observations of SN~2014J to 1D $\mch$ models with a deflagration-to-detonation transition in Sections~\ref{sec:analysis} and \ref{sec:cen-den}, in order to show the compatibility of the spherical $\mch$ explosion scenario with the observed spectra.
This type of model has provided an exceptional match to observations \citep{hoeflich17}.

\citet{srivastav16} observed SN~2014J in the optical at several epochs in the nebular phase.
They measure a low [\ion{Fe}{3}]/[\ion{Fe}{2}] ratio and suggest that this may be due to clumpiness in the ejecta.
Their data also show changing velocities for several of the spectral features.
We will look for evidence of changing velocities for a well-isolated emission line in the NIR in Section~\ref{sec:asymm}.

Overall, SN~2014J has been shown to be a fairly standard SN Ia \citep{galbany16}, which makes it ideal for probing deeper into the underlying physics of the progenitor scenarios and explosions.
There are a variety of light curve parameters for SN~2014J depending on the source article.
We use values from \citet{marion15} as our baseline, with $\Delta m_{15}(B)=1.11\pm0.02\unit{mag}$ and $m_B({\rm max})=11.68\pm0.01\unit{mag}$.
We note that unusual extinction values have been found by several analyses \citep{amanullah14,ashall14} that are significantly lower than $R_V$ for the Milky Way and potentially indicate the presence of circumstellar dust around the SN \citep{foley14b}.


\begin{figure*}[ht]
\centering
\includegraphics[width=5in]{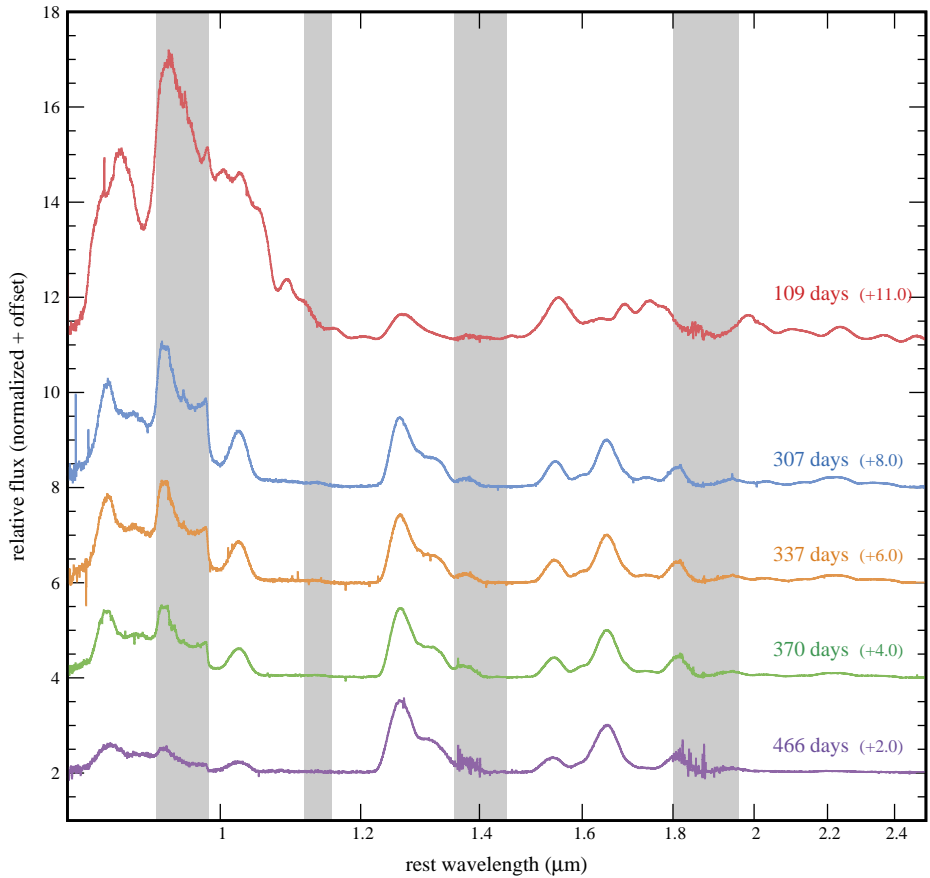}
\caption{
NIR time-series of SN~2014J extending from $109$ to $466\unit{days}$ past the estimated time of explosion. The $109\unit{day}$ observation was presented in \citet{sand16} and is shown here as an example of a spectrum during the transition phase from photospheric to nebular.
It has been normalized to the feature at $1.54\unit{\mu m}$ and then offset as indicated.
Fluxes for the other four spectra were normalized relative to the [\ion{Fe}{2}] feature at $1.65\unit{\mu m}$ and then offset as indicated.
The gray bands mark regions where the correction function is not well-behaved either because of order overlap in the cross-dispersed spectra or strong telluric absorption.\label{fig:sci_all}}
\end{figure*}

\begin{center}
\begin{deluxetable}{cchhl}
\centering
\tablecaption{SN~2014J NIR Spectroscopy\label{tab:nir_spec}}
\tablecolumns{5}
\tablewidth{0in}
\tablehead{
\colhead{\parbox[c]{0.15\textwidth}{\vspace{5pt}\centering JD Observation (2,450,000+)}} & 
\colhead{\parbox[c]{0.1\textwidth}{\vspace{5pt}\centering Epoch\tablenotemark{a} (days)}} & 
\nocolhead{\parbox[c]{0.1\textwidth}{\vspace{5pt}\centering Airmass\tablenotemark{b}}} & 
\nocolhead{\parbox[c]{0.1\textwidth}{\vspace{5pt}\centering I. Time (s)}} & 
\colhead{\parbox[c]{0.15\textwidth}{\vspace{5pt}\centering Standard Name and Stellar Type}}
\vspace{5pt}}
\startdata
\sidehead{GN-2014A-Q-8\tablenotemark{c} (PI: Sand)}
6782.18  &  109  &  1.550  &  960  &  HIP 52478 (A0III)\\
\sidehead{GN-2014B-Q-13 (PI: Diamond)}
6979.40  &  307  &  1.589  &  1440  &  HIP 32549 (A0V)\\
7009.32  &  337  &  1.596  &  1680  &  HIP 32549\\
7042.42  &  370  &  1.550  &  2400  &  HIP 32549\\
\sidehead{GN-2015A-FT-3 (PI: van Kerkwijk)}
7137.84  &  466  &  1.609  &  2160  &  HIP 50685 (A7Vn)
\enddata
\tablenotetext{a}{Using 2014 January 14.75 UT as the explosion date.}
\tablenotetext{c}{Included to show the transitional phase spectrum.}
\end{deluxetable}
\end{center}

\section{Observations and data reduction}
\label{sec:obs-red}

The observations presented in this work build on the NIR spectroscopy of SN~2014J from \citet{sand16}, which cover the photospheric phase of the SN and the transition into nebular phase, spanning $33-110\unit{days}$ post-explosion.
Three epochs of cross-dispersed NIR spectra of SN~2014J were taken using GNIRS on Gemini North.
Using the \citet{zheng14} first-light date as proxy for the explosion, these observations are at $307$, $337$, and $370\unit{days}$ after the explosion.
Data from $466\unit{days}$ post-explosion from an additional program in the Gemini Public Archives has also been included.
A summary of the late-time observations analyzed in this work is shown in Table~\ref{tab:nir_spec}.

The data were reduced using the standard procedure for Gemini observations using the \textsc{gemini} and \textsc{gnirs} modules in PyRAF.
For each of the observations, a telluric standard star was chosen to remove absorptions from our atmosphere and the detector response function in the data.
The standard for each observation was chosen to be close in angular separation and similar in airmass as the SN in order to minimize variations of atmospheric conditions between the SN and the standard.
The standard stars used for telluric corrections were HIP 52478 ($109\unit{day}$), HIP 32549 ($307$, $337$, and $371\unit{days}$), and HIP 50685 ($466\unit{day}$).
The correction functions applied to the SN data use the telluric spectrum and a synthetic spectrum from the Castelli-Kurucz Atlas \citep[ckp00\_9500 and ckp00\_7750;][]{castelli04}, where absorption lines in each were removed prior to comparison.
The correction function from the telluric standard is applied to the SN data in order to remove the effects of our atmosphere and the detector response curve.
Finally, the spectra are shifted into rest-frame wavelength assuming a heliocentric velocity for M82 of $203\unit{km}\unit{s^{-1}}$ \citep{devaucouleurs91}. 

The 1D extracted and calibrated spectra are shown in Figure~\ref{fig:sci_all}.
The spectra have been normalized to the peak of the feature at $1.54\unit{\mu m}$ for the $109\unit{day}$ spectrum, and to the peak of the $1.65\unit{\mu m}$ feature for the $307$, $337$, $370$, and $466\unit{day}$ spectra, respectively, and then offset in relative flux in order to easily compare the line profiles in each of the spectra.
The usable range of the data covers $0.82-2.5\unit{\mu m}$, however several regions of the spectra are obfuscated due to telluric features or detector sensitivity at the edges of the cross-dispersed orders; these are identified as gray regions in the figures.
It is also worth noting that most of the overlapping regions between the cross-dispersed orders fall in one of the strong telluric regions or in a region with close to zero flux, allowing some ambiguity in flux between orders and further necessitating comparisons of relative flux rather than absolute flux.
Using the GNIRS configuration for these observations, the resolving power is $R\sim1{,}800$, which corresponds to approximately $\Delta v=160\unit{km}\unit{s^{-1}}$ or $\Delta\lambda=9\unit{\AA}$ at the central wavelength of $1.65\unit{\mu m}$ for the setup.


\begin{figure}[ht]
\centering
\gridline{\fig{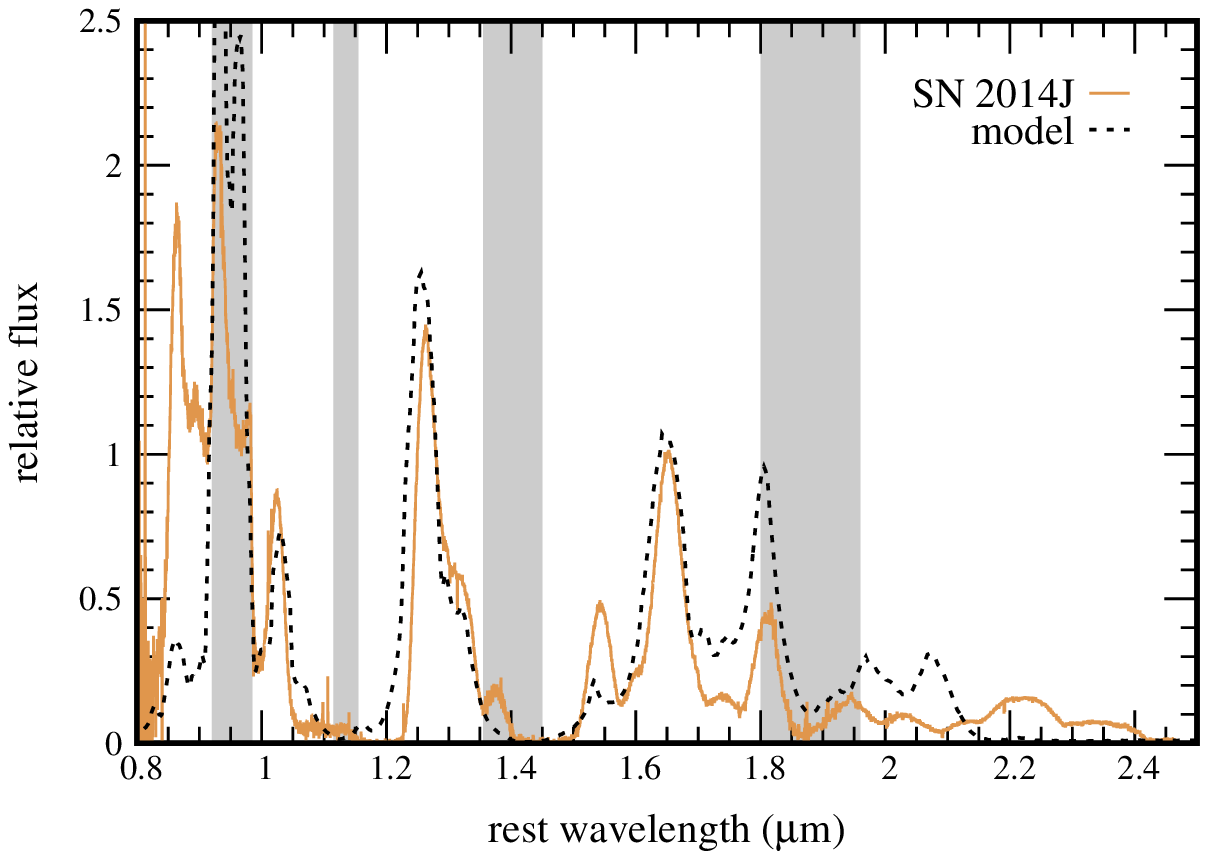}{0.45\textwidth}{(a) SN~2014J and model at $337\unit{days}$}\fig{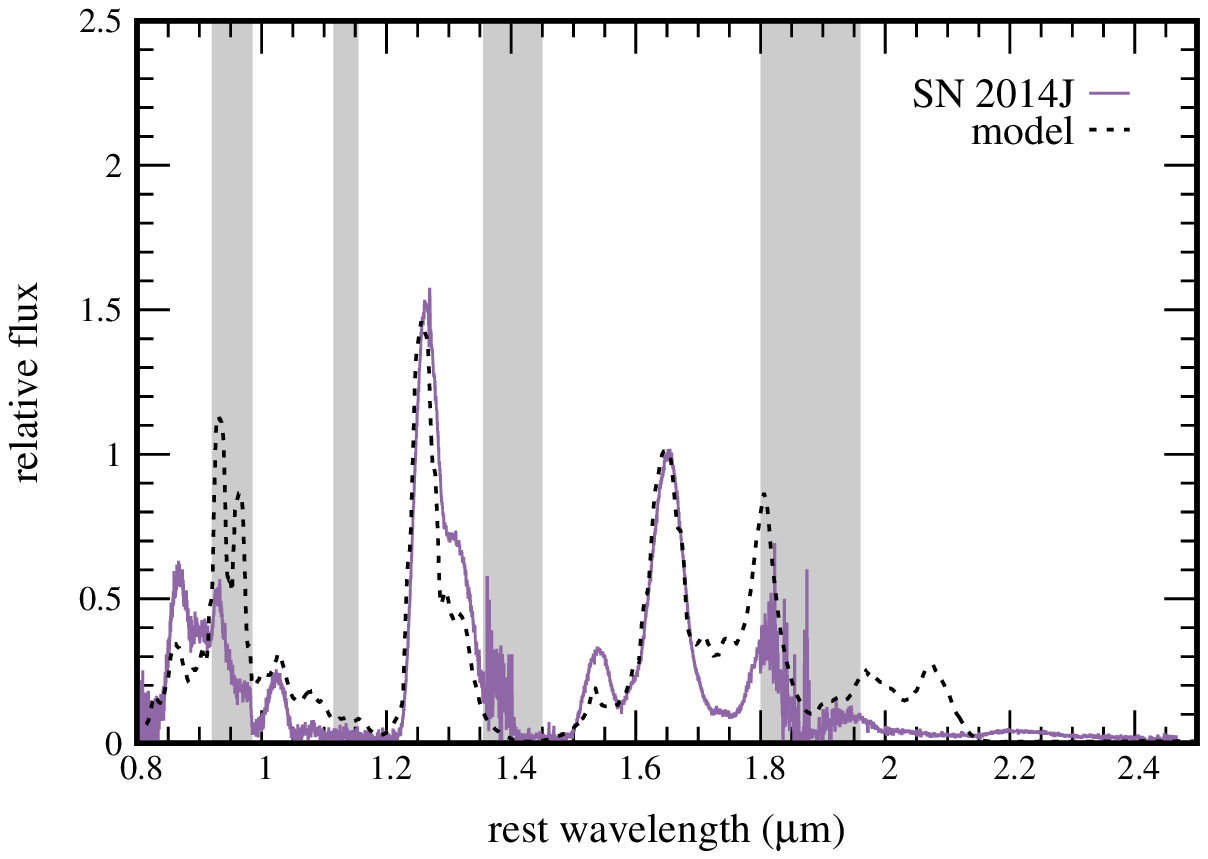}{0.45\textwidth}{(b) SN~2014J and model at $466\unit{days}$}}
\caption{
The full wavelength range observed for SN~2014J (solid lines) is shown in addition to a synthetic spectrum from the spherical $\mch$ explosion model (dotted lines) with central density $\rho_{\rm c}=0.7\times10^9\unit{g}\unit{cm^{-3}}$ and $B=10^6\unit{G}$ at $337\unit{days}$ (a) and $466\unit{days}$ (b).
The spectra are normalized to the peak of the $1.65\unit{\mu m}$ feature.
As in Figure~\ref{fig:sci_all} the gray bands mark regions where the flux is less well-known due to strong telluric contamination or order overlap.
Most of the features are due to excitations of iron group elements.
The evolution can be understood as a result of $\ce{Co}\rightarrow\ce{Fe}$ decay.
The choice of central density is optimized based on the analysis of line width of the emission feature at $1.65\unit{\mu m}$, described in detail in Section~\ref{sec:comp}.
To produce this larger wavelength spectrum as opposed to the presentation in Section~\ref{sec:comp}, we have utilized more superlevels in order to reduce the computational time, since they consist of many individual levels handled as one.
However this treatment does not produce as close of a spectral match to all of the features.\label{fig:comp-full}}
\end{figure}

\section{Analysis}
\label{sec:analysis}

We will compare the observations of SN~2014J to spherical models of delayed detonation $\mch$ explosions using emission features in the NIR spectrum.
A broad comparison of observed and model spectra is presented in Section~\ref{sec:comp}.
We use the strong $1.6440\unit{\mu m}$ [\ion{Fe}{2}] emission line as our primary comparison of individual features, as in \citet{diamond15}.
Analysis of this feature is used to probe the central density of the WD just prior to the explosion and is presented in Section~\ref{sec:cen-den}.
We also use the emission features neighboring this primary line to indicate mixing of the burning products in Section~\ref{sec:mixing}.
The symmetry and evolution of the primary [\ion{Fe}{2}] emission line is used to explore the presence of magnetic fields and asymmetries in the burning products or ignition location in Sections~\ref{sec:B-fields} and \ref{sec:asymm}.
In addition, two emission features corresponding to a blend of [\ion{S}{2}] emission lines at approximately $1.03\unit{\mu m}$ and a blend of [\ion{S}{1}], [\ion{Si}{1}], and iron group elements at approximately $1.08\unit{\mu m}$ can be used to investigate the presence and strength of magnetic fields in the ejecta in Section~\ref{sec:B-fields}.

\begin{figure}[ht]
\centering
\includegraphics[width=0.45\textwidth]{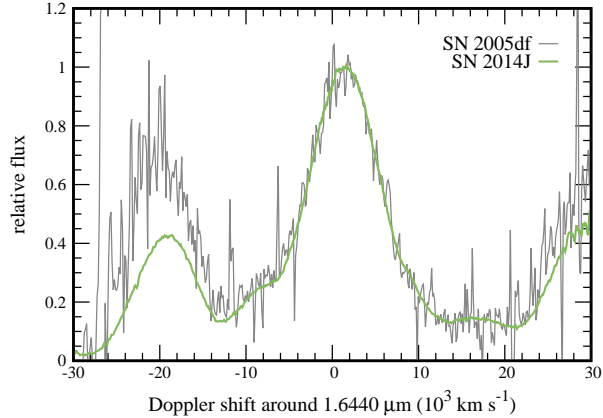}
\caption{
The region around the [\ion{Fe}{2}] emission line at $1.6440\unit{\mu m}$ is shown in velocity space for both SN~2005df \citep{diamond15} and SN~2014J at $380$ and $370\unit{days}$ post-explosion, respectively.
The SN~2005df spectrum has been shifted by $+800\unit{km}\unit{s^{-1}}$ so that the line peaks approximately match up for the sake of visual comparison.
The emission feature around $-20{,}000\unit{km}\unit{s^{-1}}$ (corresponding to approximately $1.54\unit{\mu m}$) has a higher flux level in SN~2005df, which could be attributed to stronger mixing in SN~2014J.
The effects of mixing and comparison to models is presented in Section~\ref{sec:mixing}.
\label{fig:05df}}
\end{figure}

\subsection{Models}
\label{sec:models}

The simulations are based on detailed time-dependent \edit1{non-local thermodynamic equilibrium (non-LTE)} models \citep[see][and references therein]{diamond15,hoeflich17} for light curves and spectra, which can reproduce many of the observed properties of SNe~Ia. 
As in previous works, for the overall spectra and evolution, we use superlevels with some 500 to 1000 levels for each ionization level of the iron group elements.  
Subsequently and for the detailed spectral profiles of [\ion{Fe}{2}], [\ion{Fe}{3}], [\ion{Co}{2}], and [\ion{Co}{3}] of the $1.65\unit{\mu m}$ region, we use the individual levels, the corresponding radiative transition probabilities \citep{hoeflich04b,diamond15}, and hydrogen-like approximation for the collisional rates.  
As an example, the overall spectral fits are given in Figure~\ref{fig:comp-full}.
\edit1{The model shows good agreement with respect to Doppler width and line strength of SN~2014J for the features around $1.0$, $1.3$, and $1.65\unit{\mu m}$.
However, there is some disagreement in line strength for features around $0.85$, $1.5$, and red-ward of $1.9\unit{\mu m}$, some of which may be due to the use of superlevels in the larger wavelength range synthetic spectra.
The analyses presented in the following sections {\it do not} use the larger-range synthetic spectra, but rather the more complete and computationally-intensive smaller wavelength range spectra.}

A method of comparison using the width and evolution of a relatively well-isolated [\ion{Fe}{2}] emission line at $1.6440\unit{\mu m}$ was presented in \citet{diamond15}.
The width of this line can indicate the amount of burning resulting in electron capture during the deflagration phase, which increases with increasing central density of the WD.
We use this method and expand on it in this analysis of SN~2014J.
Both similarities and differences can be seen in the observed spectra and light curves from the optical to the MIR between SNe~2014J and 2005df when using the observations presented here in addition to \citet{diamond15} and \citet{telesco15}, and a comparison of the two SNe is shown in Figure~\ref{fig:05df}.
The high signal-to-noise (S/N) of the SN~2014J data as well as a better defined continuum provide a tool to analyze beyond the presentation in \citet{diamond15} of SN~2005df.   
Within delayed detonation models of $\mch$ explosions, these similarities can be attributed to a similar amount of deflagration burning between the two SNe. 
Therefore, we use spherical delayed-detonation models with the same basic parameters from \citet{diamond15} as a baseline.
Because of the evidence of limited mixing we will take multi-dimensional effects such as deflagration mixing as perturbations to the baseline. 
As noted in \citet{diamond15}, mixing will have an effect similar to a reduced central density.


\begin{figure*}[ht]
\centering
\gridline{\fig{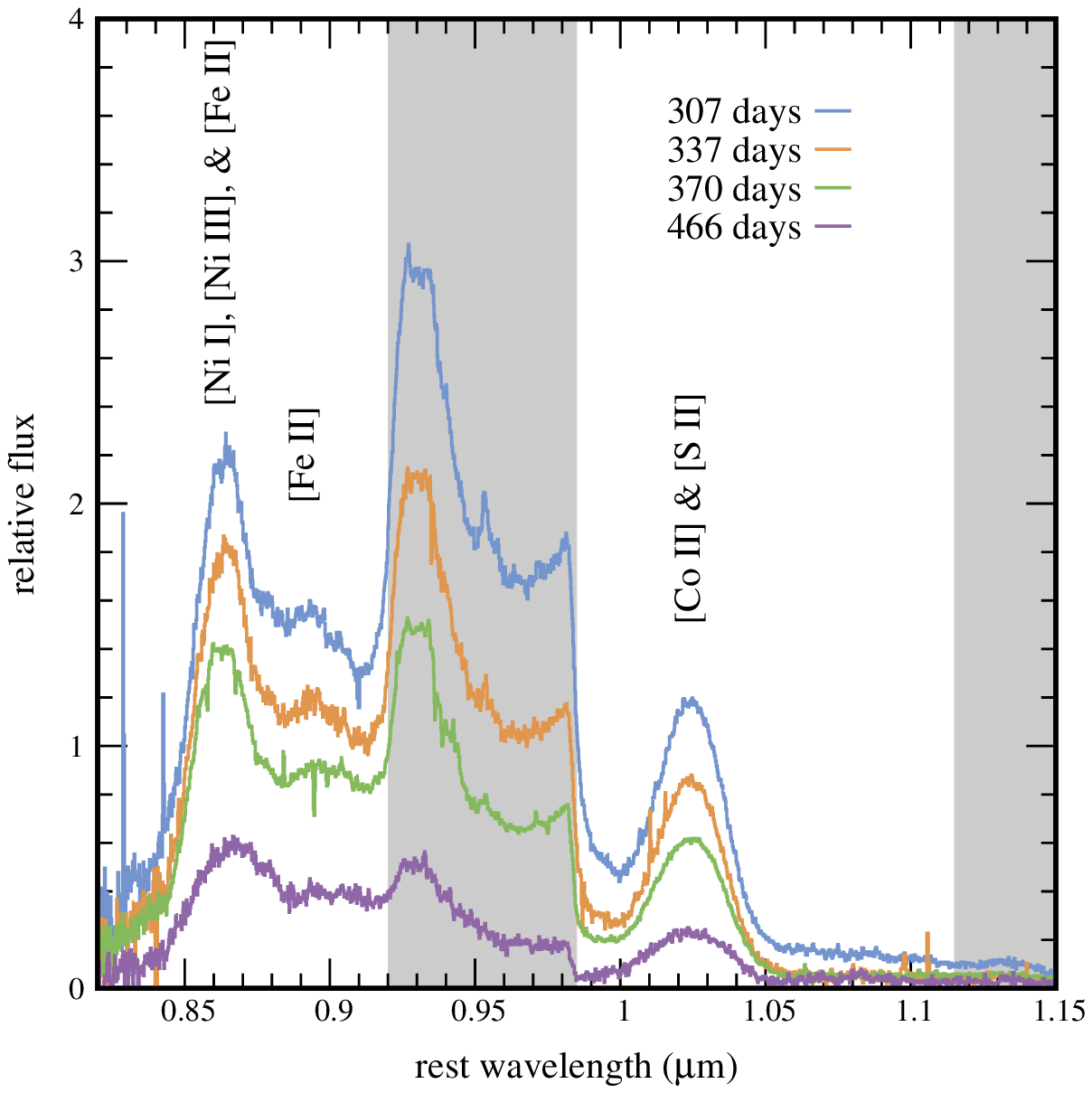}{0.45\textwidth}{(a)}
\fig{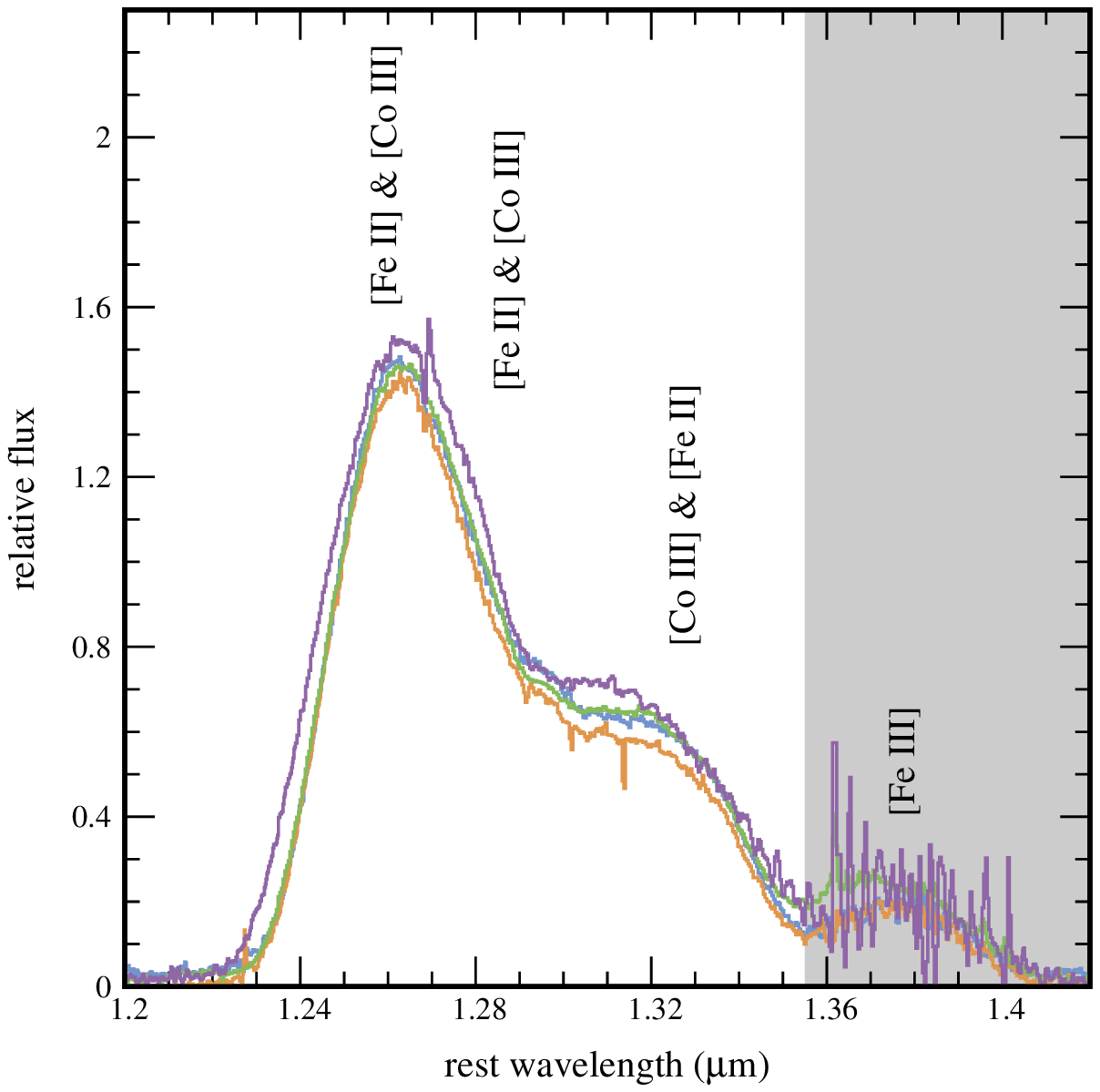}{0.45\textwidth}{(b) corresponding roughly to the $J$ band}}
\gridline{\fig{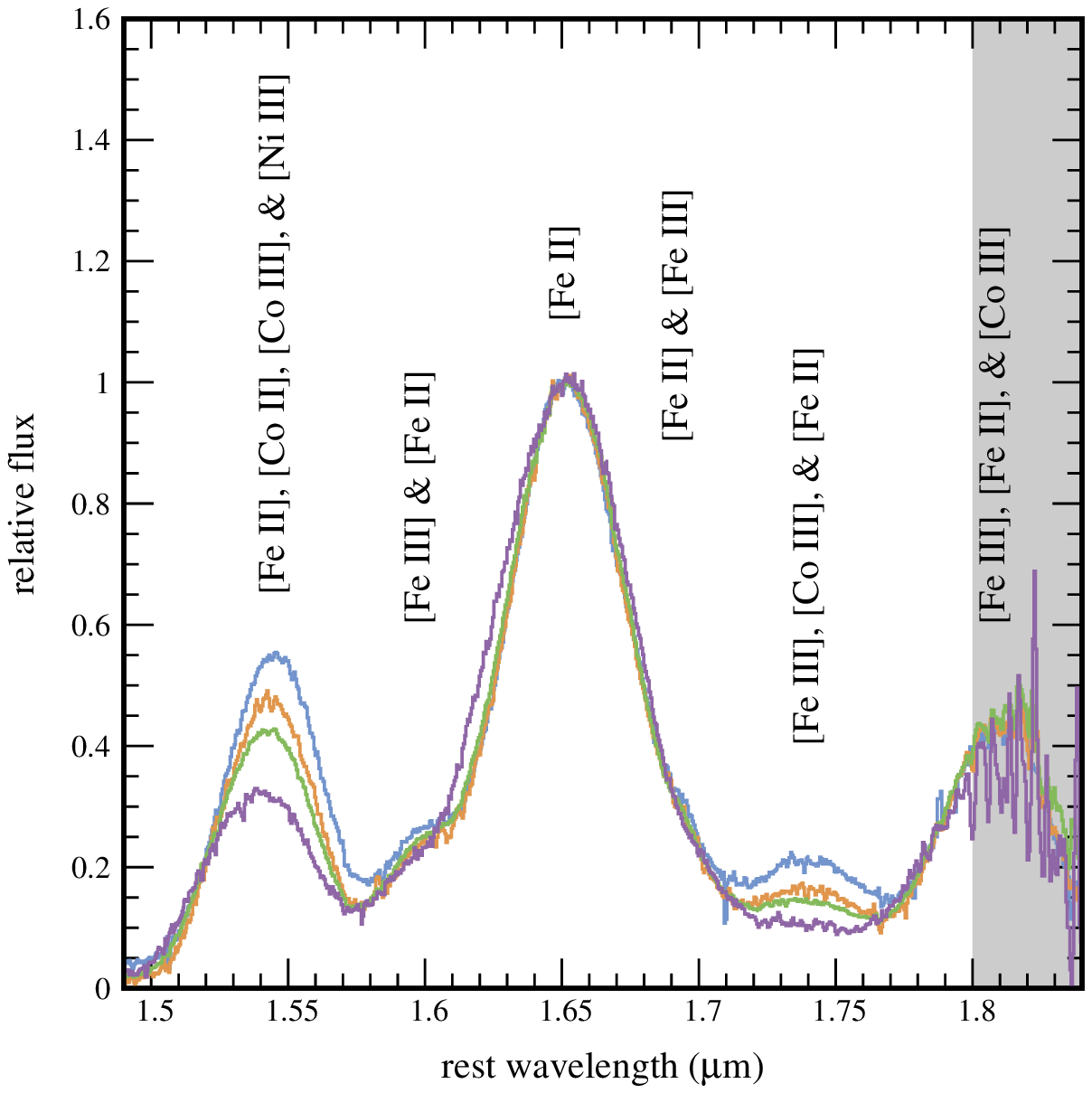}{0.45\textwidth}{(c) corresponding roughly to the $H$ band}
\fig{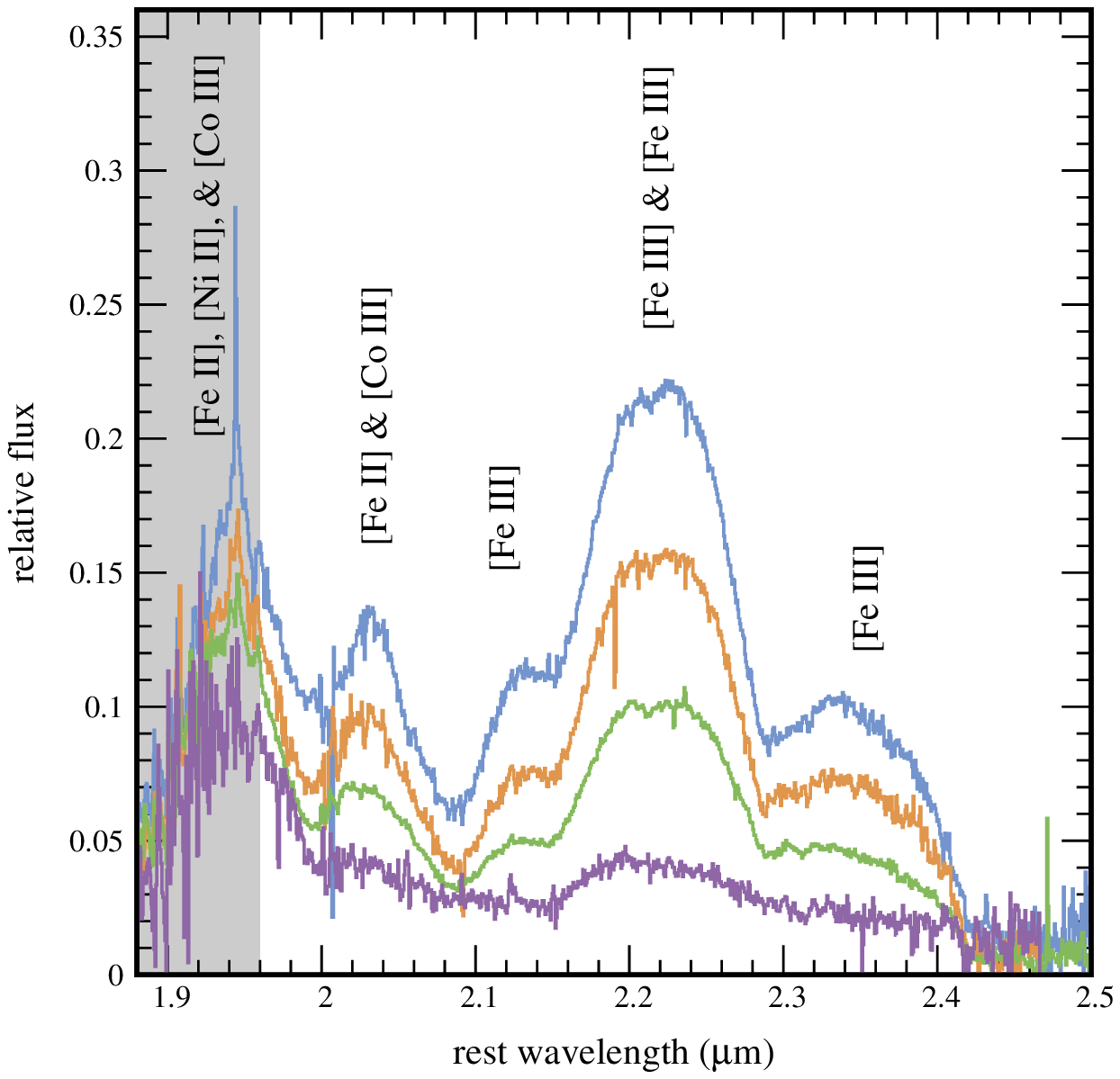}{0.45\textwidth}{(d) corresponding roughly to the $K$ band}}
\caption{
The evolution of the SN~2014J spectrum is shown for the four epochs analyzed in this work.
Fluxes for the SN were normalized so that the feature at $1.65\unit{\mu m}$ is $1$ in all of the epochs.
The regions in gray are strongly affected by telluric absorption and order overlap.
Forbidden emission lines of iron and cobalt dominate the spectra and show the expected evolution from \ce{{}^{56}Ni} in the decay of $\ce{Co}\rightarrow\ce{Fe}$.
Line species that are the main contributors to the observed emission features, based on modeling and comparison to \citet{bowers97}, are labeled here and detailed in Table~\ref{tab:line-id}.\label{fig:evol}}
\end{figure*}

\subsection{Comparison of the observations and the models}
\label{sec:comp}

As described in Section~\ref{sec:models}, the SN~2014J data are compared to synthetic spectra from spherical $\mch$ explosion models with a detonation following a deflagration phase.
As in \citet{diamond15}, the free parameters for the models are the amount of burning prior to the \edit1{delayed detonation transition (DDT, }parameterized as a transition density, $\rho_{\rm tr}$), the central density ($\rho_c$) of the WD prior to the explosion, the zero-age main sequence mass of the progenitor and the metallicity of the progenitor \citep{hoeflich98a}.
Like in \citet{diamond15}, the initial $\rho_c$ of the different models are $0.5$, $0.7$, $0.9$, $1.1$, $2.0$, and $4.0\times10^9\unit{g}\unit{cm^{-3}}$.
The $\rho_{\rm tr}$ of the models is set to maintain similar outer layers of burning products between the various models.

\edit1{Full mixing of the inner region provides peaked rather than broad profiles \citep{hoeflich04b}, and is not favored by abundance mapping using tomography \citep{mazzali00,ashall14,ashall18}.
Additionally, however, even limited mixing of the burning products will show an effect of ionization on the line strengths of different emission features in the spectrum.}
As an example of the mixing effect, in this work we use models both with no mixing and with limited mixing.
As a template for the latter case we applied a mixing scale height of $2{,}000\unit{km}\unit{s^{-1}}$, based on the chemical distribution observed in the SNR S Andromeda \citep{fesen15,fesen17}. 
Stable \ce{Ni}, \ce{Co}, and \ce{Fe} material is mixed out and radioactive \ce{Fe} and \ce{Co} material is mixed inwards towards higher density regions, changing the ratio between the blended [\ion{Co}{3}]-[\ion{Fe}{2}] feature at $1.54\unit{\mu m}$ relative to the main [\ion{Fe}{2}] emission feature at $1.65\unit{\mu m}$. 
Emission from higher densities implies more `quasi-continuum' and less emission in the $1.54\unit{\mu m}$ feature. 

Using the emission lines that show the strongest contributions in synthetic spectra from the $\mch$ models, features from the SN~2014J data are identified in Figure~\ref{fig:evol} and Table~\ref{tab:line-id}.
Most of the features are blends of forbidden emission lines from iron group elements, specifically iron, cobalt, and nickel.
There are also several strong sulfur lines, especially in the heavily-blended region on the blue side of $1.1\unit{\mu m}$.
These identifications are comparable to the emission lines found in the nebular NIR spectra of SN~2005df \citep{diamond15} and in analyses of other nebular SNe Ia spectra \citep{bowers97,spyromilio04}.
As in \citet{diamond15}, the [\ion{Fe}{2}] emission line at $1.6440\unit{\mu m}$ is the only feature in the NIR that does not include significant blending from neighboring lines and thus is ideal for an individual study.

The regions with the most significant evolution of the line profiles are $0.82-1.05\unit{\mu m}$ (Figure~\ref{fig:evol}a) and $1.9-2.5\unit{\mu m}$ (Figure~\ref{fig:evol}d).
The heavily blended \ce{Fe}-\ce{Co} feature between $1.23-1.35\unit{\mu m}$ in Figure~\ref{fig:evol}b shows some evolution between the different epochs, however some of this may be attributed to uncertainties in the order matching for the cross-dispersed spectra and the use of a different standard star for the $466\unit{day}$ observation.

\begin{figure}[ht]
\centering
\includegraphics[width=0.45\textwidth]{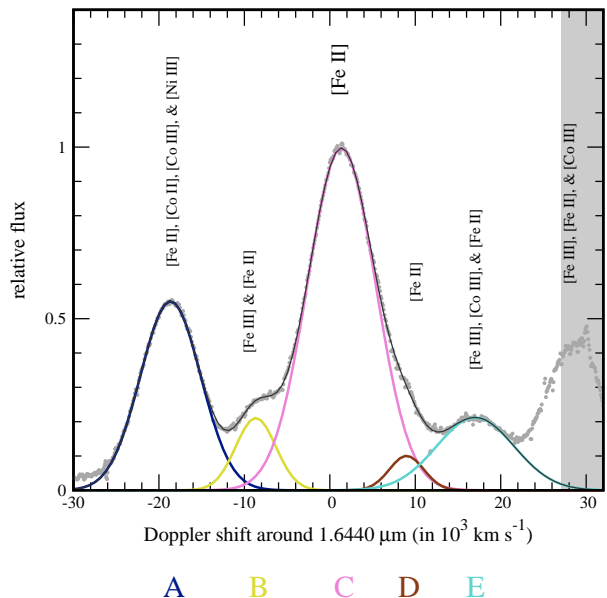}
\caption{
The region around $1.6440\unit{\mu m}$ is shown for SN~2014J at \edit1{$307\unit{days}$} (gray points).
The spectrum is broken into five Gaussian components, labeled Feature A, B, C, D, and E for easy identification in the text.
The central feature (Line C, \edit1{pink}) is composed of a single strong emission line, corresponding to [\ion{Fe}{2}].
There are satellite lines that weakly contribute to the wings of the central emission line, corresponding to blends of [\ion{Fe}{2}] and [\ion{Fe}{3}], and neighboring lines sporting a mix of iron group burning products.
See Table~\ref{tab:line-id} for the species, transition, and rest wavelength of individual lines that contribute based on comparison with the synthetic spectrum.
\label{fig:line-composition}}
\end{figure}

Focusing on the two strongest features in the $H$-band region (Figure~\ref{fig:evol}c) we use a Gaussian to fit the central regions of each feature in velocity space to get an idea of the evolution of the line width and peak offset for these emission features.
Figure~\ref{fig:line-composition} shows a fit of five Gaussian profiles for the $H$-band region.
Depending on the region of the emission feature chosen, there is an uncertainty based on using this Gaussian fitting method of approximately $\pm40\unit{km}\unit{s^{-1}}$, which translates to $\approx2\unit{\AA}$ for both of these strongest features in $H$ and is about $4$ times smaller than the resolving power of the detector setup (approximately $160\unit{km}\unit{s^{-1}}$).
Table~\ref{tab:vel} shows the location of the peaks and line widths of these two main $H$-band features at the various epochs.

\begin{figure}[ht]
\centering
\includegraphics[width=0.45\textwidth]{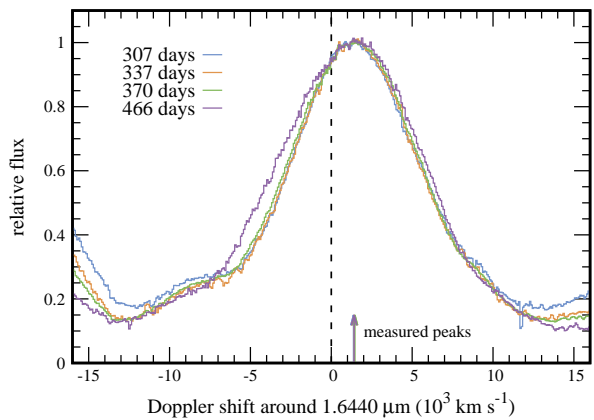}
\caption{
Evolution of the observed $1.65\unit{\mu m}$ feature, identified as the $1.6440\unit{\mu m}$ [\ion{Fe}{2}] emission line.
Note that the measured peak of the feature is shifted to the red of the expected $0\unit{km}\unit{s^{-1}}$ line (shown by the dotted black line).
See Section~\ref{sec:asymm} for a discussion of the relative velocity shifts for each of the epochs.\label{fig:vel1644}}
\end{figure}

\begin{figure}[ht]
\centering
\includegraphics[width=0.45\textwidth]{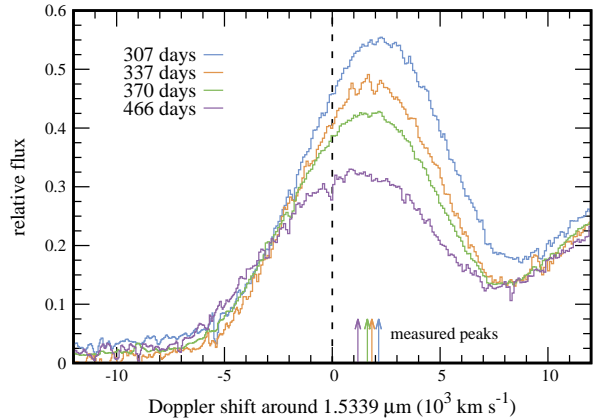}
\caption{
Evolution of the observed $1.54\unit{\mu m}$ feature, identified as a blend of the $1.5339\unit{\mu m}$ [\ion{Fe}{2}], $1.5474\unit{\mu m}$ [\ion{Co}{2}], and $1.5488\unit{\mu m}$ [\ion{Co}{3}] emission lines.
Note that the measured peak of the feature is not constant for all four epochs, which is consistent with the decay of the cobalt moving the blended line center closer to the iron line.
See Section~\ref{sec:asymm} for a discussion of the relative velocity shifts for each of the epochs.\label{fig:vel15339}}
\end{figure}

We would caution against using blended lines to probe the velocities of individual species.
It is not trivial to analyze the separate emission lines for multiple reasons, including: 
\begin{enumerate*}[label=\itshape\alph*\upshape)]
\item the line strengths are sensitive to changes in ionization, 
\item the line strengths and widths depend on the details of mixing, 
\item there are uncertainties in the atomic data for the line transitions, and 
\item wavelength offsets from the expected values will depend on asymmetries in the distribution of burning products or transition to detonation.\end{enumerate*}
Line A ($\approx1.54\unit{\mu m}$ in the data) is a poor choice for further analysis because it is a blend of four different atomic transitions, each significantly contributing to the feature in the synthetic spectra.
Many emission features throughout the optical and infrared are heavily blended and may lead to incorrect conclusions.
In contrast, Line C ($\approx1.65\unit{\mu m}$) is not expected to be strongly affected for the reasons listed above since one emission line is the dominant contributor and the satellite features are from the same atomic transition (see Table~\ref{tab:line-id}) and only contribute to the flux at a very low level, as seen in Figure~\ref{fig:line-composition}.
Figures~\ref{fig:vel1644} and \ref{fig:vel15339} show the nebular SN~2014J spectra in velocity space around the $1.6440\unit{\mu m}$ [\ion{Fe}{2}] and $1.5339\unit{\mu m}$ [\ion{Fe}{2}] lines, respectively.
Both emission features peak red-ward of the expected line center based on the rest frame of M82, and we will discuss this in more detail below.
In velocity space, these line profiles should be symmetric around the peak if there are no asymmetries in the distribution of burning products and the transition to detonation was not off-center.

\startlongtable
\begin{center}
\begin{deluxetable}{ccccc}
\centering
\tablecaption{Line Identification for SN~2014J\label{tab:line-id}}
\tablecolumns{5}
\tablewidth{0in}
\tablehead{
\colhead{\parbox[c]{0.025\textwidth}{\vspace{4pt}\centering $\lambda_{\rm rest}$ $(\unit{\mu m})$}} & 
\colhead{\parbox[c]{0.075\textwidth}{\vspace{4pt}\centering Contr. to $\lambda$ $(\unit{\mu m})$}\vspace{4pt}} & 
\colhead{Species} & 
\colhead{Transition} & 
\colhead{$J-J'$}
}
\startdata
\sidehead{Corresponding to Figure~\ref{fig:evol}a:}
0.8469 & 0.861 & [\ion{Ni}{1}] & \ce{{}^3D-{}^1D} & $1-2$\\
0.8502 & 0.861 & [\ion{Ni}{3}] & \ce{{}^3F-{}^1D} & $2-2$\\
0.8619 & 0.861 & [\ion{Fe}{2}] & \ce{a {}^4F-a {}^4P} & $9/2-5/2$\\
0.8894 & 0.890 & [\ion{Fe}{2}] & \ce{a {}^4F-a {}^4P} & $7/2-3/2$\\
1.0240\tablenotemark{*} & 1.023 & [\ion{Co}{2}] & \ce{a {}^3F - b {}^3F} & \nodata\\
1.0331\tablenotemark{*} & 1.023 & [\ion{S}{2}] & \ce{{}^2D^o-{}^2P^o} & \nodata\\
\sidehead{Corresponding to Figure~\ref{fig:evol}b:}
1.2567 & 1.259 & [\ion{Fe}{2}] & \ce{a {}^6D - a {}^4D} & $9/2-7/2$\\
1.2728 & 1.259 & [\ion{Co}{3}] & \ce{a {}^4P - a {}^2D_2} & $5/2-5/2$\\
1.2946 & 1.289 & [\ion{Fe}{2}] & \ce{a {}^6D - a {}^4D} & $5/2-5/2$\\
1.3105 & 1.289 & [\ion{Co}{3}] & \ce{a {}^4P - a {}^2D_2} & $3/2-5/2$\\
1.3105 & 1.327 & [\ion{Co}{3}] & \ce{a {}^4P - a {}^2D_2} & $3/2-5/2$\\
1.3209 & 1.327 & [\ion{Fe}{2}] & \ce{a {}^6D - a {}^4D} & $7/2-7/2$\\
1.4098 & 1.377 & [\ion{Fe}{3}] & \ce{{}^1D_4 - {}^1F} & $2-3$ \\
\sidehead{Corresponding to Figure~\ref{fig:evol}c:}
1.5339 & 1.544 & [\ion{Fe}{2}] & \ce{a {}^4F - a {}^4D} & $9/2-5/2$\\
1.5474 & 1.544 & [\ion{Co}{2}] & \ce{a {}^5F - b {}^3F} & $5-4$\\
1.5488 & 1.544 & [\ion{Co}{3}] & \ce{a {}^2G - a {}^2H} & $9/2-9/2$\\
1.5511 & 1.544 & [\ion{Ni}{3}] & \ce{{}^3P - {}^1G} & $2-4$\\
1.5875 & 1.597 & [\ion{Fe}{3}] & \ce{{}^3G - {}^3D} & $5-3$\\
1.5999 & 1.597 & [\ion{Fe}{2}] & \ce{a {}^4F - a {}^4D} & $7/2-3/2$\\
1.6440\tablenotemark{a} & 1.650 & [\ion{Fe}{2}] & \ce{a {}^4F - a {}^4D} & $9/2-7/2$\\
1.6843\tablenotemark{*} & 1.691 & [\ion{Fe}{2}] & \ce{a {}^4F - a {}^4D} & \nodata\\
1.7110\tablenotemark{*} & 1.691, 1.739 & [\ion{Fe}{3}] & \ce{{}^3G - {}^3D} & \nodata\\
1.7413 & 1.739 & [\ion{Co}{3}] & \ce{a {}^2G - a {}^2H} & $9/2-11/2$\\
1.7454 & 1.739 & [\ion{Fe}{2}] & \ce{a {}^4F - a {}^4D} & $3/2-1/2$\\
1.7643 & 1.739 & [\ion{Co}{3}] & \ce{a {}^2F - a {}^2H} & $7/2-11/2$\\
1.7926\tablenotemark{*} & 1.807 & [\ion{Fe}{3}] & \ce{{}^3G - {}^3D} & \nodata\\
1.8027\tablenotemark{*} & 1.807 & [\ion{Fe}{2}] & \ce{a {}^4F - a {}^4D} & \nodata\\
1.8028 & 1.807 & [\ion{Fe}{2}] & \ce{a {}^2H - b {}^2H} & $9/2-9/2$\\
1.8129\tablenotemark{*} & 1.807 & [\ion{Fe}{2}] & \ce{a {}^4D - a {}^4P} & \nodata\\
1.8214 & 1.807 & [\ion{Co}{3}] & \ce{a {}^4P - a {}^2P} & $3/2-1/2$ \\
\sidehead{Corresponding to Figure~\ref{fig:evol}d:}
1.9141 & 1.942 & [\ion{Fe}{2}] & \ce{a {}^4D - a {}^4P} & $3/2-1/2$\\
1.9393 & 1.942 & [\ion{Ni}{2}] & \ce{{}^4F - {}^2F} & $9/2-7/2$\\
1.9581 & 1.942 & [\ion{Co}{3}] & \ce{a {}^4P - a {}^2P} & $1/2-1/2$\\
1.9675 & 2.034 & [\ion{Fe}{2}] & \ce{a {}^4D - a {}^4P} & $5/2-5/2$\\
2.0028 & 2.034 & [\ion{Co}{3}] & \ce{a {}^4P - a {}^2P} & $5/2-3/2$\\
2.1457 & 2.117 & [\ion{Fe}{3}] & \ce{a {}^3H - a {}^3G} & $4-3$\\
2.2184 & 2.217 & [\ion{Fe}{3}] & \ce{a {}^3H - a {}^3G} & $6-5$\\
2.2427 & 2.217 & [\ion{Fe}{3}] & \ce{a {}^3H - a {}^3G} & $4-4$\\
2.3485 & 2.353 & [\ion{Fe}{3}] & \ce{a {}^3H - a {}^3G} & $5-5$
\enddata
\tablenotetext{a}{Primary line used in this analysis.}
\tablenotetext{*}{Multiple $J-J'$ values contribute.} 
\end{deluxetable}
\end{center}

Another interesting region to mention in the spectra corresponds to the $K$ band (Figure~\ref{fig:evol}d).
High S/N $K$-band spectra during the nebular phase are rare for SNe Ia.
The $K$-band data allow for the identification of emission lines (see Table~\ref{tab:line-id}) after comparison with synthetic spectra from the aforementioned models as well as from \citet{bowers97}.
\edit1{Potential contributions from stable nickel lie in and around the obscured telluric region.}
As in the other NIR regions, iron and cobalt dominate in this wavelength region, and the features evolve as expected based on the decay of \ce{{}^{56}Ni} to \ce{{}^{56}Co} to \ce{{}^{56}Fe}.

We identify the $1.65\unit{\mu m}$ feature (Figure~\ref{fig:line-composition}, Line~C) as predominantly [\ion{Fe}{2}] with some small contributions in the wings from neighboring [\ion{Fe}{2}] and [\ion{Fe}{3}] emission lines in the blue wing (Line~B) and red wing (Line~D). 
The location of the peak of the $1.65\unit{\mu m}$ feature is shown in Figure~\ref{fig:vel1644} and is fairly consistent between all four epochs: $1.652\unit{\mu m}$.
Any differences can be attributed to the uncertainties in the method of determining the line peak at the $\unit{\AA}$ level, as described above, in combination with the resolving power of the detector at the $10\unit{\AA}$ level.
This peak offset corresponds to a Doppler shift of approximately $+1{,}400\unit{km}\unit{s^{-1}}$ and is much larger than the stellar velocity, $\approx+110\unit{km}\unit{s^{-1}}$, measured at this position in the host M82 \citep{greco12} and the velocity range of the strong \ion{Na}{1} D absorption, $\approx+40$ to $+250\unit{km}\unit{s^{-1}}$, observed in the SN spectrum during the photospheric phase \citep{goobar14,welty14,foley14b}.
Figure~\ref{fig:vel1644} shows no noticeable change of the $1.65\unit{\mu m}$ emission feature in the first three epochs.
Evolution is noticeable in the neighboring features, outside of the $\pm10{,}000\unit{km}\unit{s^{-1}}$ range.
However, there is a noticeable widening of the line profile in the $466\unit{day}$ spectrum, particularly in the blue wing.
The line width of the $1.65\unit{\mu m}$ feature will be compared with predictions from modeling in Section~\ref{sec:cen-den}.

\begin{center}
\begin{deluxetable}{cccc}
\centering
\tablecaption{Measured Peaks of the $1.65\unit{\mu m}$ and $1.54\unit{\mu m}$ Features and Comparison to Known Emission Lines\label{tab:vel}} 
\tablecolumns{4}
\tablewidth{0in}
\tablehead{
\colhead{\parbox[c]{0.1\textwidth}{\vspace{4pt}\centering Epoch\tablenotemark{a} (days)}\vspace{2pt}} & 
\colhead{\parbox[c]{0.1\textwidth}{\vspace{4pt}\centering Central $\lambda$ $(\unit{\mu m})$}} & 
\colhead{\parbox[c]{0.1\textwidth}{\vspace{4pt}\centering Vel. Shift $(\unit{km}\unit{s^{-1}})$}} & 
\colhead{\parbox[c]{0.1\textwidth}{\vspace{4pt}\centering $\sim$Width $(\sigma)$ ($\unit{km}\unit{s^{-1}}$\,)}}
}
\startdata
\sidehead{\parbox[c]{7cm}{The $1.65\unit{\mu m}$ feature relative to $1.6440\unit{\mu m}$ [\ion{Fe}{2}].}}
307 & 1.6515 & 1{,}360 & 3{,}980\\
337 & 1.6516 & 1{,}410 & 4{,}040\\
370 & 1.6513 & 1{,}330 & 4{,}150\\
466 & 1.6518 & 1{,}420 & 4{,}450\\
\sidehead{\parbox[c]{7cm}{The $1.54\unit{\mu m}$ feature relative to $1.5339\unit{\mu m}$ [\ion{Fe}{2}].\tablenotemark{b}}}
307 & 1.5449 & 2{,}150 & 3{,}580\\
337 & 1.5433 & 1{,}840 & 3{,}510\\
370 & 1.5422 & 1{,}630 & 3{,}650\\
466 & 1.5400 & 1{,}190 & 4{,}140
\enddata
\tablenotetext{a}{Using 2014 January 14.75 UT \citep{zheng14} as the explosion date.}
\tablenotetext{b}{See the discussion in Sections~\ref{sec:obs-red} and \ref{sec:asymm} about the effect of line blending.}
\end{deluxetable}
\end{center}

We identify the $1.54\unit{\mu m}$ feature (Figure~\ref{fig:line-composition}, Line A) as a blend of an [\ion{Fe}{2}], a [\ion{Co}{2}], and a [\ion{Co}{3}] emission line at $1.5339$, $1.5474$, and $1.5488\unit{\mu m}$, respectively.
There is also potentially some contribution from [\ion{Ni}{3}] at $1.5511\unit{\mu m}$ from stable nickel produced through electron capture in the early stages of the explosion.
Unlike the $1.65\unit{\mu m}$ feature, the $1.54\unit{\mu m}$ feature shows significant evolution in the offset of the measured peak of the line profile, the relative strength of the line, and the shape of the line profile (Figure~\ref{fig:vel15339}).
Based on the rest wavelength of each of the contributing emission lines, the peak of the feature should shift blue-ward as radioactive \ce{{}^{56}Co} decays to stable \ce{{}^{56}Fe}, which is exactly what is seen in the observed spectra, with the location of the measured peak ranging from $1.5449$ to $1.5400\unit{\mu m}$ (see Table~\ref{tab:vel}).
However, even when iron dominates the emission feature, the line center is still redshifted when compared to the host rest-frame, which is consistent with the results for the $1.65\unit{\mu m}$ feature.
The line strength relative to the $1.65\unit{\mu m}$ feature decreases as time goes on, which is also indicative of the decreasing amount of cobalt in this blended feature.


\begin{figure}[ht]
\centering
\includegraphics[width=0.45\textwidth]{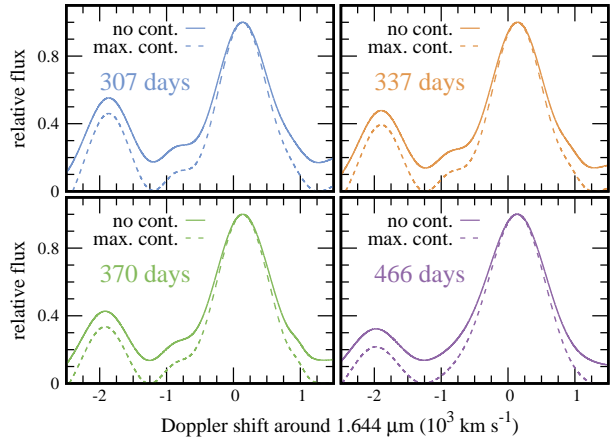}
\caption{
The smoothed SN~2014J spectrum is shown normalized to the $1.65\unit{\mu m}$ emission feature assuming no continuum and a maximum continuum for each epoch observed.
Based on the choice of continuum level, the pseudo line width varies for a chosen relative flux height, $\myF$.\label{fig:comp-continuum}}
\end{figure}

\begin{figure}[ht]
\centering
\includegraphics[width=0.45\textwidth]{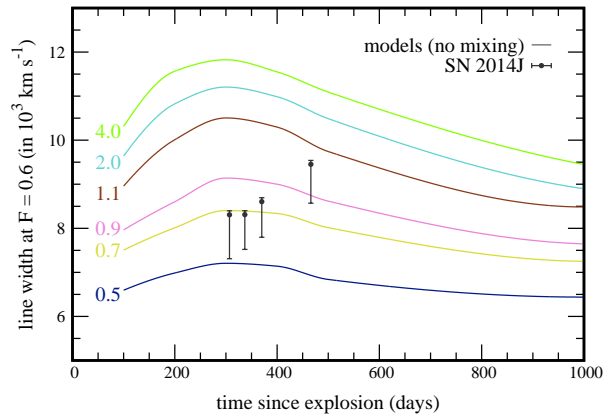}
\caption{
The line width taken at a relative flux height of $\myF=0.6$ for the spherical DDT models, without mixing, are shown for a range of central densities: $\rho_{c9}=0.5$ to $4.0$ (in units of $10^9\unit{g}\unit{cm^{-3}}$, labels shown to the left of each curve) and for SN~2014J (black points).
The lower error-bars on the data points include continuum subtraction.
The early epochs of the observed SN~2014J spectra are consistent with a $\rho_c$ around $0.7\times10^9\unit{g}\unit{cm^{-3}}$.
The measured line widths at other $\myF$ values are also in agreement with the results shown here. \label{fig:LW-comp}}
\end{figure}

\section{Results \& discussion}
\label{sec:results}

\subsection{Effects of central density}
\label{sec:cen-den}

In the spherical delayed-detonation models of $\mch$ WD explosions discussed in Section~\ref{sec:analysis}, the $\rho_c$ of the WD just prior to the thermonuclear runaway determines whether or not there will be a significant amount of electron capture in the central regions during the deflagration, producing stable iron group elements such as \ce{{}^{58}Ni}.
The well-isolated [\ion{Fe}{2}] emission line at $1.6440\unit{\mu m}$ can be used as a probe for this ``radioactive hole'' since the extent and geometry of the emitting region will determine the shape and width of the feature.

\edit1{Although some earlier spherical DDT models with unmixed electron capture regions produced almost boxy, ``flat-topped'' line profiles \citep{nomoto84,hoeflich04b}, updated electron capture rates do not produce as large amounts of stable iron group elements \citep{brachwitz00,langanke00,hoeflich06a}.
Hence, a less severe shape in the synthetic spectra is seen for central densities of the WD even up to $4\times10^9\unit{g}\unit{cm^{-3}}$.
During the phase from $150-200\unit{days}$ after the explosion, both $\gamma$-photons and positrons will contribute to the energy deposition in the ejecta; however, after about $300\unit{days}$, positrons will be the dominant energy source \citep{penney14}.
The effect on line profile of a central ``hole'' in the radioactive burning products caused by electron capture can be seen in \citet[][Figure 13]{stritzinger15}, effectively widening the emission feature so there is no sharply-peaked profile but not producing a ``flat-top'' -- a misnomer.
The old electron capture rates, such as in W7, will over-predict the ratio of \ce{Ni}/\ce{Fe} produced in the SN Ia, and may affect analyses that rely on the predictive power of this ratio, as in \citet{maguire18}.
}

Two methods of measuring a pseudo-width for the feature are chosen: 
\begin{enumerate}[nosep]
\item[a)] the width at a chosen relative flux, $\myF$, where the maximum of the $1.65\unit{\mu m}$ feature is normalized to 1 and we have assumed no continuum flux; and
\item[b)] the width at a chosen relative flux, $\myF$, where the maximum of the $1.65\unit{\mu m}$ feature is normalized to 1 and we have assumed a {\em maximum} value for \edit1{a flat continuum flux} based on the minimum value on the red-ward side of the emission feature between $1.7-1.75\unit{\mu m}$.
\end{enumerate}
Note that method (b) is an over-subtraction and merely demonstrates the extreme end of the pseudo-width since there are minor emission features contributing in that region, not only quasi-continuum.
As an example of this, Figure~\ref{fig:comp-continuum} demonstrates the effect of continuum on the pseudo line width of the $1.65\unit{\mu m}$ emission feature using both methods -- the SN~2014J spectra have been smoothed in order to do this.
This question of continuum subtraction can be incorporated into an uncertainty on the pseudo-width and will produce a narrower line for a chosen $\myF$.
\edit1{As $\myF$ decreases, the effect of the continuum flux on the line width increases, correspondingly increasing the error-bars on the data.
The $\myF=0.6$ line width plot maximized the separation of the models while also minimizing confusion from the continuum flux.
We note that $\myF=0.5$ corresponds to the more-commonly quoted full width at half maximum (FWHM) value.
For $\myF=0.6$, the line widths using method (a) versus (b) introduces an error of up to $1{,}000\unit{km}\unit{s^{-1}}$ and varies slightly depending on the epoch.}

Figure~\ref{fig:LW-comp} shows the pseudo-widths of the observations and models with several central densities for a chosen relative flux height \edit1{of $\myF=0.6$}.
The data points indicate method (a) {\em no continuum} and the lower error-bar \edit1{extended down to incorporate} method (b) {\em maximum continuum}.
The choice of $\myF=0.6$ to measure the line width of the feature is fairly arbitrary, however plots from $\myF=0.5-0.8$ are consistent with each other.
Below about $\myF=0.4$, the neighboring emission lines begin to influence the evolution of the line width and are more sensitive to uncertainties in abundances and mixing in the SN~Ia and in the transition strengths of these forbidden lines.
Above about $\myF=0.9$, noise in the observations begins to strongly influence the measured line width.
The sensitivity of the pseudo-width on the background continuum becomes apparent in Figs.~\ref{fig:comp-continuum} and \ref{fig:LW-comp}, where we compare the line width observed for SN~2014J and the theoretical models. 
The observed line width for SN~2014J increases approximately linearly with time. 
In principle, an increase may be attributed to the positron transport effect, however an exponential increase with time rather than linear would be expected based on this. 
The first two epochs (\edit1{$307$} and \edit1{$337\unit{days}$}) are consistent with a $\rho_c$ of $0.7\times10^9\unit{g}\unit{cm^{-3}}$, and models with mixing do not significantly affect the results.
However, the later epochs (\edit1{$370$} and \edit1{$466\unit{days}$}) do not follow any of the line width trends of the models considered -- these spectra and their deviations from the models will be probed further in Section~\ref{sec:asymm}.

For a $\mch$ model, a $\rho_c$ of $0.7\times10^9\unit{g}\unit{cm^{-3}}$ is quite low. 
This implies a high accretion rate where the progenitor is not able to increase much in $\rho_c$ prior to ignition.
The upper limit for the accretion rate is given by the Eddington limit, and the material being accreted matters because violent shell flashes will not allow excess material to build up on the WD.
With the $\rho_c$ found, the data are consistent with a SD progenitor scenario if the dominant mass-transfer channel is \ce{He}-accretion rather than \ce{H}-accretion from the companion star.
Accretion of \co\, from a degenerate companion would also be possible \citep{nomoto82a,hoeflich98b,maeda16}.


\begin{figure}[ht]
\centering
\includegraphics[width=0.45\textwidth]{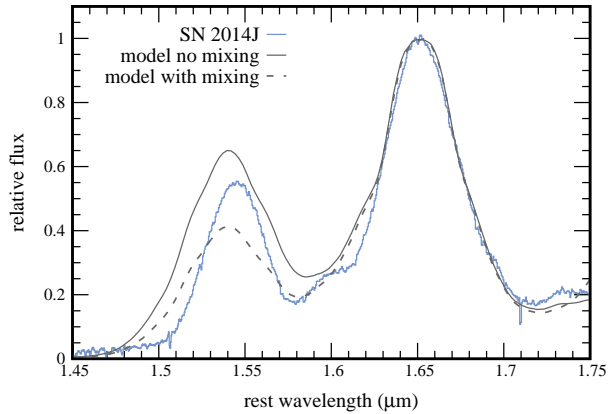}
\caption{
A comparison of the SN~2014J data at $307\unit{days}$ and the $\rho_c=0.7\times10^9\unit{g}\unit{cm^{-3}}$ models, both with and without mixing.\label{fig:mixing-effect}}
\end{figure}

\subsection{Mixing of the chemical distribution}
\label{sec:mixing}

It is natural to expect mixing of the burning products during the deflagration phase due to Rayleigh-Taylor instabilities, and indeed, mixing of the chemical layers is seen in most 2D and 3D models.
In spherical models, such as the ones used in this analysis, mixing must be added \edit1{by hand} after the deflagration structure has been calculated and before calculating the spectrum.
However, large-scale mixing is not favored because of the radially-structured observations of SNe Ia \citep[see][Figure 14 and corresponding discussion]{hoeflich02}.
A possible physical reason for the suppression of mixing is the confinement of positrons due to strong magnetic fields in the ejecta \citep{hristov18}.

\edit1{Unlike full or large-scale mixing, limited mixing in the central regions} does not significantly alter the line width of the $1.6440\unit{\mu m}$ [\ion{Fe}{2}] emission feature.
However, the amount of mixing can be probed by comparing the various emission features, since mixing will affect the ionization in the ejecta. 
The $1.65\unit{\mu m}$ feature is composed of mainly [\ion{Fe}{2}] and the $1.54\unit{\mu m}$ feature is a blend of iron and cobalt, as discussed in Section~\ref{sec:analysis}.
\edit1{The limited mixing introduced in the models used in this analysis moves some of the radioactive iron group burning products inwards to higher density regions.
With the increase in density, the average collisions and recombination rates increase, favoring lower ionization states.
The relative height of the two features at a given epoch will depend on the mixing, with an increase in mixing suppressing the $1.54\unit{\mu m}$ feature in the synthetic spectra.}
Comparison with the observed profiles in SN~2014J (shown in Figure~\ref{fig:mixing-effect}) suggests some mixing, although a smaller amount than implied in the synthetic spectra based on the chemical distribution seen in S Andromeda \citep{fesen15,fesen17}.


\begin{figure}[ht]
\centering
\includegraphics[width=0.45\textwidth]{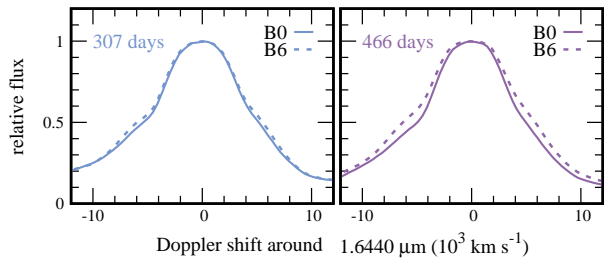}
\caption{
The effect of magnetic fields in the ejecta can be seen in the evolution of the [\ion{Fe}{2}] line profile.
The models with central density of $\rho_c=0.7\times10^9\unit{g}\unit{cm^{-3}}$ with field strengths of $B=0\unit{G}$ (solid lines) and $B=10^9\unit{G}$ (dotted lines) are shown at two different epochs.
The magnetic field effect is more evident as the ejecta expand.\label{fig:bs}}
\end{figure}

\begin{figure}[ht]
\centering
\includegraphics[width=0.45\textwidth]{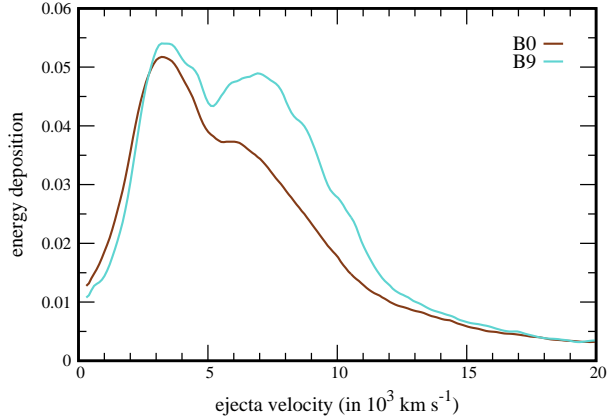}
\caption{
At late times, the energy deposition is dominated by positrons and transport effects become important at the outer layers \citep{penney14}. 
The energy deposition corresponds to the initial \ce{{}^{56}Ni} distribution and magnetic fields present in the ejecta. 
Its outer edge marks the transition to layers of incomplete burning, namely \ce{Si} and \ce{S}-rich layers, and the inner layers mark the inner region of electron-capture iron-group elements \citep{hoeflich02}.  \label{fig:energy-dep}}
\end{figure}

\begin{figure}[ht]
\centering
\gridline{\fig{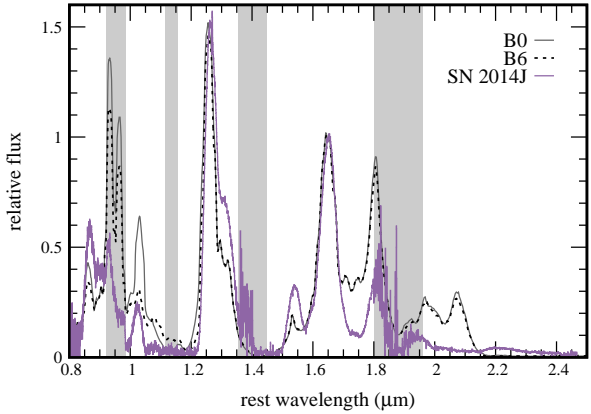}{0.45\textwidth}{(a) \edit1{The inner, lower velocity layers are hardly affected by the presence of magnetic fields, resulting in similar line profiles for burning products near the center of the WD and the overall spectra at late-times.}}\fig{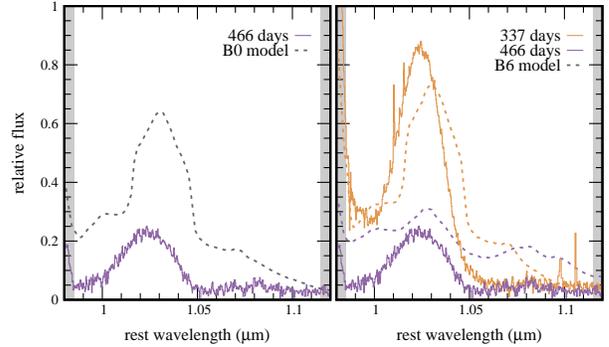}{0.45\textwidth}{(b) Zoomed in on the region around $1.05\unit{\mu m}$, which shows the largest effect of the embedded magnetic field on the synthetic spectrum.}}
\caption{
Comparison of SN~2014J and models with and without $B$ fields.
Synthetic spectra come from models with $\rho_c=0.7\times10^9\unit{g}\unit{cm^{-3}}$ and $B=0$ and $B=10^6\unit{G}$. 
For $B<10^6\unit{G}$, positrons escape the NSE region and we see a relative increase in ionization and heating of the quasi-equilibrium (QSE) region.
A low magnetic field boosts the emission in [\ion{S}{2}] at $1.03\unit{\mu m}$ compared to [\ion{S}{1}] at $1.08\unit{\mu m}$\edit1{, over-predicting the flux at $1.03\unit{\mu m}$ by a factor of three.
The [\ion{S}{2}] emission feature is shifted around $1{,}600\unit{km}\unit{s^{-1}}$ blue-ward in SN~2014J compared to the synthetic spectra.
\label{fig:comp}}}
\end{figure}

\subsection{Magnetic fields}
\label{sec:B-fields}

Magnetic fields start to play a role or have an effect in the evolution of the spectrum after about three hundred days, which becomes even more visible as time passes, as can start to be seen in Figure~\ref{fig:bs}.
Magnetic fields will inhibit positron flow, which results in line profiles remaining broad later than expected from $B$-free models.
In the case of SN~2014J, the emission feature at $1.65\unit{\mu m}$ not only remains broad through all of our observations, but actually increases in width, as seen in Figure~\ref{fig:vel1644}.
Some of this may be attributed to very strong magnetic fields, especially if they are turbulent in nature, however additional factors may also be at work and are discussed in the next section.

With the full wavelength models, we note other potential NIR features that can be used to probe magnetic fields.
A feature emerges at $\approx1.033\unit{\mu m}$ that is mostly due to a strong blend of [\ion{S}{2}] at $1.0323$, $1.0339$, and $1.0373\unit{\mu m}$ and a feature at $\approx1.08\unit{\mu m}$ produced by [\ion{S}{1}], [\ion{Si}{1}], [\ion{Fe}{2}], [\ion{Fe}{3}], and [\ion{Ni}{2}]. 
Looking at the energy deposition in Figure~\ref{fig:energy-dep}, at low magnetic field strength, there is less positron trapping and local deposition of energy.
The positron escape out from the central regions and in turn heat the regions of incomplete burning (at around $10{,}500\unit{km}\unit{s^{-1}}$), boosting emission for [\ion{S}{2}] and reducing emission for [\ion{S}{1}].
\edit1{This results in a flux too large by a factor of $3\times$ for the $1.03\unit{\mu m}$ emission feature, hence our argument that a non-zero magnetic field is indicated.
We note that in both the $B=0\unit{G}$ (B0) and $B=10^6\unit{G}$ (B6) models, the flux in the $1.08\unit{\mu m}$ emission feature is over-predicted.}

Since most of the emission features in the NIR are due to transitions of iron group elements, the overall NIR line profile looks very similar for the synthetic spectra with and without an embedded magnetic field (Figure~\ref{fig:comp}a).
As can be seen in Figure~\ref{fig:comp}b, though, SN~2014J does not have a discernible emission feature at $1.08\unit{\mu m}$ and the emission feature at $1.03\unit{\mu m}$ has a similar flux level to the $B=10^6\unit{G}$ model.
However, because the flux levels between the different cross-dispersed orders is somewhat uncertain in part due to strong telluric absorption \edit1{in the region $1.115\unit{\mu m}<\lambda<1.15\unit{\mu m}$}, we cannot put a hard limit on the strength of the magnetic field embedded in the SN ejecta.


\begin{figure}[ht]
\centering
\gridline{\fig{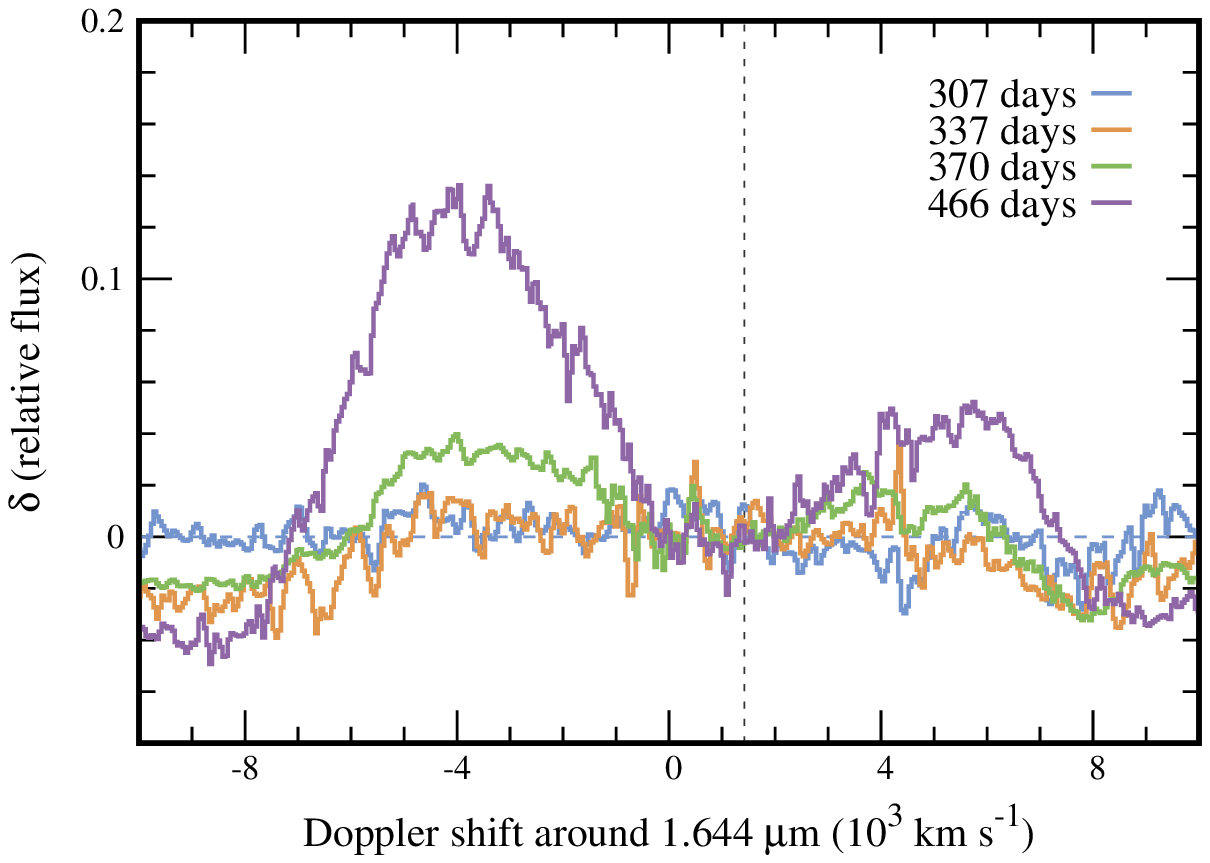}{0.45\textwidth}{(a) difference assuming no continuum subtraction}\fig{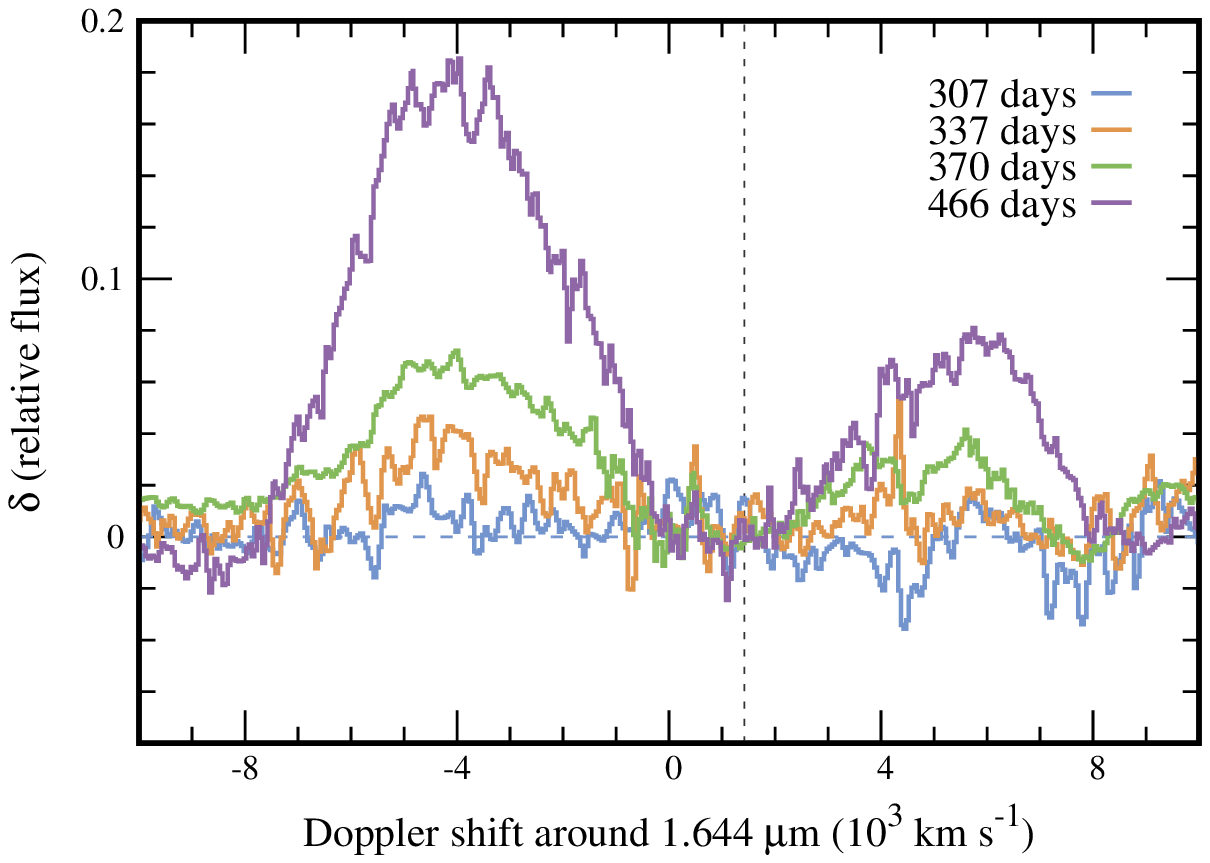}{0.45\textwidth}{(b) difference assuming maximum continuum subtraction}}
\caption{
Differences between the relative flux of the $1.65\unit{\mu m}$ emission feature for individual epochs and the smoothed $307\unit{day}$ observation.
The black dotted line shows the approximate line center, based on the Gaussian fit from Section~\ref{sec:comp} and Table~\ref{tab:vel}.
In (a) no continuum has been subtracted from the spectra.
In (b) a maximum value for the continuum at each epoch has been subtracted from the spectra.
Regardless of the choice of continuum, there is an obvious asymmetry present in the $466\unit{day}$ spectrum.\label{fig:diff}}
\end{figure}

\begin{figure}[ht]
\centering
\includegraphics[width=0.435\textwidth]{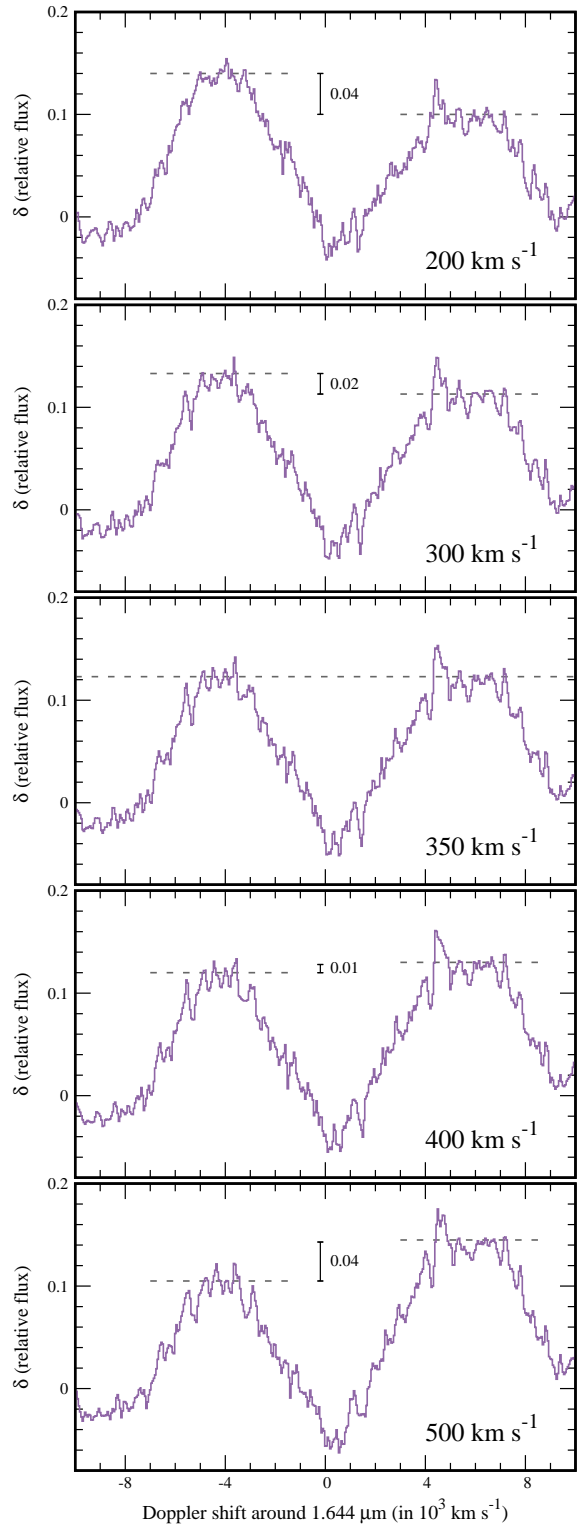}
\caption{
The $466\unit{day}$ spectrum, assuming maximum continuum subtraction, has been shifted in velocity-space when compared to the $307\unit{day}$ spectrum before the difference $\delta$ was taken to determine the asymmetry in the central region of the burning products.
\edit1{Approximate vertical peak separation is shown in each panel, with the $350\unit{km}\unit{s^{-1}}$ offset producing approximately level peaks.}\label{fig:diff-off}}
\end{figure}

\subsection{Asymmetries in the chemical distribution}
\label{sec:asymm}

Any asymmetries seen in the spectra may be due to a combination of an off-center ignition and an off-center transition to detonation in a $\mch$ explosion with a DDT \citep{hoeflich02c,zingale05}.
We will note that some of the alternative explosion scenarios may also produce similar asymmetries in the late-time spectra \citep{pakmor10,pakmor11,pakmor12}.


\subsubsection{Increasing line width of the $1.65\unit{\mu m}$ emission feature}
A widening of the $1.65\unit{\mu m}$ line profile can be seen in the $\sigma$-value found in the Gaussian fit from Section~\ref{sec:comp} and Table~\ref{tab:vel}, around $4{,}500\unit{km}\unit{s^{-1}}$ compared to $4{,}000\unit{km}\unit{s^{-1}}$ for previous epochs.
Likewise, if we consider Figure~\ref{fig:LW-comp}, the jump in approximately $1{,}000\unit{km}\unit{s^{-1}}$ in the pseudo line width from the expected trend if the data followed the $\rho_c=0.7\times10^9\unit{g}\unit{cm^{-3}}$ model is suggestive of an asymmetry in the explosion.
The pseudo line widths at $\myF=0.6$ are $8{,}308\unit{km}\unit{s^{-1}}$, $8{,}310\unit{km}\unit{s^{-1}}$, $8{,}605\unit{km}\unit{s^{-1}}$, and $9{,}453\unit{km}\unit{s^{-1}}$ from $307-466\unit{days}$, respectively.
Using the $0.7\times10^9\unit{g}\unit{cm^{-3}}$ model from Section~\ref{sec:cen-den}, the mass fraction corresponding to each of these velocities increases from about $0.52$ for $307$ and $337\unit{days}$, to $0.55$ for $371\unit{days}$, to $0.63$ for $466\unit{days}$.

These mass fractions do not describe the central regions of the ejecta, so this asymmetry is unlikely to be caused by an off-center ignition. 
The transition from deflagration to detonation occurs in one location and the front takes time to wrap around the exploding WD.
This would produce a difference in the ionization structure because the far side would continue to expand and decrease in temperature and density prior to the transition to detonation, producing an asymmetric distribution of the outer \ce{{}^{56}Ni} \citep{hoeflich06b,fesen07}.
As a result, the average velocity of energy input is shifted towards higher velocities, and depending on the orientation of the embedded magnetic field, will result in a broader line or a strong shift of the line.
In order to have this time-delay, the positrons must be kept relatively local, which is another indicator that high magnetic fields are needed.
Note that the Larmor radius must be significantly smaller than the size of the \ce{{}^{56}Ni} region, and following \citet{penney14}, this translates to a lower limit for $B$ of $10^5\unit{G}$.\footnote{Off-center ionization is currently being studied at FSU (Alec Fisher, PhD thesis in preparation).}

\subsubsection{A small shift in the central wavelength of the $1.65\unit{\mu m}$ emission feature}
We also know from Figure~\ref{fig:vel1644} that the line itself seems to shift blue-ward in addition to the increasing width.
However, looking at the measured peaks of the emission feature, the central wavelength shift is not significant since it is on the same order as the uncertainty ($\sim100\unit{km}\unit{s^{-1}}$).  
In order to probe this additional asymmetry, we will look at differences in relative flux between the epochs, which are very sensitive to small changes.
By looking at the difference of two spectra, $\delta$, the development of asymmetries in the line profiles especially stands out.
However, the difference plots are also dependent on the chosen continuum of the spectrum for each epoch.
Since the spectra have all been normalized to the [\ion{Fe}{2}] emission line, the difference is zero at approximately the line-center.
The difference in spectrum with respect to the smoothed $307\unit{day}$ spectrum is shown in Figure~\ref{fig:diff}, assuming no continuum subtraction (Figure~\ref{fig:diff}a) and removing and subtracting a `maximum' continuum for each epoch prior to differencing (Figure~\ref{fig:diff}b), as in Section~\ref{sec:cen-den}.

In Figure~\ref{fig:diff} we show the distribution of the differential profiles as a function of time.
\edit1{For the interpretation of this asymmetry, we note that there are secondary components from species in the wings of the $1.65\unit{\mu m}$ feature \citep[see][Figure~4]{hoeflich04b}. 
The non-[\ion{Fe}{2}] components will result in an uneven distribution of the differentials of about a factor of 1.12 to 1.29 at 300 and 500 days, respectively \citep{hengeler14}. 
However, these components lie well out in the wings past $\pm10{,}000\unit{km}\unit{s^{-1}}$, and we do not take them into account when analyzing the asymmetry here.}

To get a size scale for this asymmetry, we have offset the $466\unit{day}$ spectrum toward the red in velocity space when compared to the $307\unit{day}$ spectrum and measure the excess flux as $\delta(\rm{rel. flux})$.
By adjusting the observed rest frame for the different epochs, we obtain an overall shift of $\approx 350\unit{km}\unit{s^{-1}}$ of the kinematic from the chemical `center'. 
The velocity-step is given in each of the subplots in Figure~\ref{fig:diff-off}.
Looking at Figure~\ref{fig:vel1644}, neighboring emission features start to contribute noticeably outside of the range $-6{,}000$ to $9{,}000\unit{km}\unit{s^{-1}}$.
The $337\unit{day}$ spectrum does not show significant evolution of the line profile when compared to the $307\unit{day}$ spectrum.
In contrast, the $370\unit{day}$ spectrum shows line broadening, and the level is fairly symmetric on the blue and red sides of line-center, although it might show hints of a slight asymmetry if the maximum continuum is removed.
The $466\unit{day}$ spectrum shows a very interesting evolution.
The red side of the line profile has broadened slightly more than in the previous epoch, however the blue side of the line has broadened dramatically, now above $\delta=0.1$.
The two offsets at $\approx350\unit{km}\unit{s^{-1}}$ most equalize the excess in the emission feature's wings.
In these two subplots a gray line shows the level of the excess, at $\delta\approx0.12$.
What would cause this asymmetric line profile evolution?

For the $\rho_c=0.7\times10^9\unit{g}\unit{cm^{-3}}$ model, the mass fraction that a velocity shift of $350\unit{km}\unit{s^{-1}}$ corresponds to is the inner $\approx 2\times10^{-4}$ of the WD.
One potential explanation based on where this asymmetry is located is an off-center ignition of the initial deflagration \citep[see][]{hoeflich06b,maeda10b}.
As time passes in the nebular phase, more and more positrons produced in regions with radioactive iron group elements are able to become non-local and light up the central regions where the density was high enough for electron capture to occur, creating stable iron group elements.

In Figure~\ref{fig:vel1644}, we see an overall shift of the emission feature line center corresponding to about $1{,}330-1{,}420\unit{km}\unit{s^{-1}}$ relative to the rest frame of the host galaxy. 
Likewise, all of the differentials shown in Figure~\ref{fig:diff} have a $\delta=0$ point around a Doppler shift of $\approx1{,}350\unit{km}\unit{s^{-1}}$ since there is no significant shift of the peak.
This is likely too large to be explained by the orbital motion of the WD in the host galaxy.
Alternatively, the offset of the line profile center may be attributed to the change of the velocity of outer \ce{{}^{56}Ni}, which is also consistent with the off-center DDT scenario \citep{hoeflich06b,fesen07,maeda10a,maeda11}.  
A shift may be related to the off-center ignition and subsequent rising of a plume or an off-center detonation that results in an global asymmetry in the \ce{{}^{56}Ni} distribution. 
Plume velocities are of the order of a few hundred kilometers per second and the deflagration phase lasts some $1-2\unit{s}$ \citep{gamezo05}.
We note that it is smaller than the several $1{,}000\unit{km}\unit{s^{-1}}$ reported by \citet{maeda12}, which assumed a large plume with an overall off-set.


\begin{figure}[ht]
\centering
\includegraphics[width=0.45\textwidth]{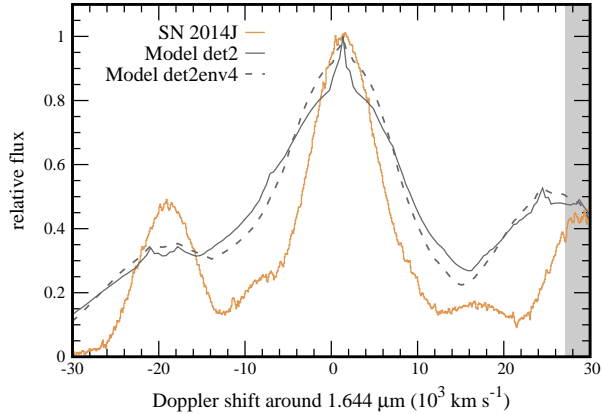}
\caption{
A comparison of the SN~2014J data at $337\unit{days}$ and two merger models at the same epoch.
These models produce iron group burning products at higher-velocity than the $\mch$ models, making the emission features wider than the observational data.\label{fig:merger}}
\end{figure}

\subsection{Alternative scenarios}
\label{sec:alt}

A comparison of the observed data with two alternative scenario models is shown in Figure~\ref{fig:merger}.
These two models include Models DET2, a ``He detonation'', and DET2ENV4, a ``classical merger'' of two WDs \citep[see][for details including parameter values, i.e. $\rho_c$, $\rho_{\rm tr}$, and resulting observables]{hoeflich96}.
Large rotations of the primary WD may be expected in the case of the classical merger, which will introduce an additional dependence on the viewing angle of the system.
In both alternative scenarios, the central densities are not high enough to produce stable nickel, so a larger amount of \ce{{}^{56}Fe} is produced compared to the $\mch$ models.
The resulting spectra are expected to have stronger low-velocity components and a larger width, producing more triangular line profiles.
The expected spectra in both alternative scenarios explored show emission lines with a central peak and much larger low-velocity contributions because of the large amount of \ce{{}^{56}Ni} produced and the distribution of burning products in the ejecta, resulting in wider characteristic line profiles.
Neither of the merger models provide a good match to the overall shape or evolution of the SN~2014J spectra.
Clearly there are many different merger models in the literature with a variety of spectral predictions and an in-depth exploration is not possible in this work.
We encourage modelers to produce synthetic spectra in the NIR and MIR in addition to the more traditional optical wavelength range.
However, the constraint from the emission line width of SN~2014J strongly disfavors some of the available progenitor channels for SNe, namely the merger and He detonation scenarios we considered.


\section*{Conclusions}
\label{sec:conclusions}

The spectra presented in this work show SN~2014J in the nebular phase at $307$, $337$, $370$, and $466\unit{days}$ post-explosion in the NIR region from approximately $0.8-2.5\unit{\mu m}$.
These observations join a relatively small sample of Type Ia supernova spectra close enough for late-time NIR follow-up.
They provide information about the very central regions of the SNe, which are obscured at early times because of optical depth effects.
We use a well-isolated [\ion{Fe}{2}] emission line at $1.6440\unit{\mu m}$ to compare with synthetic models and find the SN~2014J data are consistent with expected line widths for a white dwarf with central density of $\rho_c=0.7\times10^9\unit{g}\unit{cm^{-3}}$ undergoing a $\mch$ explosion with a deflagration-to-detonation transition, a set of models that has historically been found to exceptionally match observations.
The alternative models included in this analysis show emission line profiles that have narrow centrally-peaked lines on top of a broad emission feature, which is not consistent with the SN~2014J observations (see Figure~\ref{fig:merger}).

Using the flux levels of the $1.6440\unit{\mu m}$ [\ion{Fe}{2}] emission line with the neighboring feature around $1.54\unit{\mu m}$, we show that there is some limited mixing of the central regions of burning products when compared to a no-mixing model. 
However, mixing similar to what is seen in the SNR S Andromeda produces too much suppression of the neighboring feature.
The limited mixing in SN~2014J may be produced in part by Rayleigh-Taylor instabilities, expected in all 3D $\mch$ models that has been partially suppressed by physics not yet included in full models, such as magnetic fields in the ejecta.
Although the SN~2014J spectra are at a better S/N than most of the previous late-time NIR data in the literature, the apparent asymmetry in the $466\unit{day}$ spectrum limits our ability to quantify magnetic field strength in the ejecta.
The asymmetry in the spectral evolution corresponds to a position about $350\unit{km}\unit{s^{-1}}$ from the kinetic center and could be the result of asymmetries in the burning products or an off-center ignition.
Emission lines at $1.03\unit{\mu m}$, a blend of [\ion{S}{2}], and $1.08\unit{\mu m}$, a blend of [\ion{S}{1}], [\ion{Si}{1}], and several transitions of iron group elements, provide an additional probe of magnetic field strength and morphology, and disfavor a $B=0\unit{G}$ scenario.
Additional high S/N late-time NIR spectra, especially when complemented by MIR observations from the upcoming James Webb Space Telescope, will help constrain progenitor systems and give insight into magnetic field strength and morphology.


\section*{Acknowledgments}
\label{sec:ack}

We would like to thank many colleagues and collaborators for their support.
This research is based on observations using Gemini North (\edit1{GN-2014A-Q-8, PI: Sand;} GN-2014B-Q-13, PI: Diamond; GN-2015A-FT-3, PI: van Kerkwijk).
T. Diamond is supported by an appointment to the NASA Postdoctoral Program at the Goddard Space Flight Center, administered by Universities Space Research Association under contract with NASA.
The work presented in this paper has been supported in part by NSF awards AST-1008343 (PI: M.~M. Phillips), AST-151764 (PI: D.~J. Sand), AST-1613426 (PI: M.~M. Phillips), AST-1613472 (PI: E.~Y. Hsiao), and AST-1715133 (PI: P. Hoeflich).
E.~Y. Hsiao also acknowledges the support provided by the Florida Space Grant Consortium.
M.~D. Stritzinger is funded by a research grant (13261) from the VILLUM FONDEN.
We would also like to express our thanks to Peter van Hoof for creation of {\it The Atomic Line List} (V2.05B18) at {\small\url{http://www.pa.uky.edu/~peter/newpage/}} and the NIST ADS Team for creation of {\it The NIST Atomic Spectra Database} (V5.2) at {\small\url{http://physics.nist.gov/asd}}.

\facility{Gemini:Gillett(GNIRS)}




\begin{thebibliography}{}
\bibitem[Amanullah et al.(2014)]{amanullah14} {Amanullah}, R., {Goobar}, A., {Johansson}, J., et al. 2014, {\apjl}, 788, L21
\bibitem[Ashall et al.(2014)]{ashall14} {Ashall}, C., {Mazzali}, P., {Bersier}, D., et al. 2014, {\mnras}, 445, 4427
\bibitem[Ashall et al.(2018)]{ashall18} {Ashall}, C., {Mazzali}, P.~A., {Stritzinger}, M.~D., et al. 2018, {\mnras}, 477, 153
\bibitem[Bowers et al.(1997)]{bowers97} {Bowers}, E.~J.~C., {Meikle}, W.~P.~S., {Geballe}, T.~R., et al. 1997, {\mnras}, 290, 663
\bibitem[Brachwitz et al.(2000)]{brachwitz00} {Brachwitz}, F., {Dean}, D.~J., {Hix}, W.~R., et al. 2000, {\apj}, 536, 934
\bibitem[Branch et al.(1995)]{branch95} {Branch}, D., {Livio}, M., {Yungelson}, L.~R., {Boffi}, F.~R., \& {Baron}, E. 1995, {\pasp}, 107, 1019
\bibitem[Burns et al.(2014)]{burns14} {Burns}, C.~R., {Stritzinger}, M., {Phillips}, M.~M., et al. 2014, {\apj}, 789, 32
\bibitem[Cao et al.(2014)]{cao14} {Cao}, Y., {Kasliwal}, M.~M., {McKay}, A., \& {Bradley}, A. 2014, {ATEL}, 5786
\bibitem[Castelli \& Kurucz(2004)]{castelli04} {Castelli}, F., \& {Kurucz}, R.~L. 2004, {Proceedings of the IAU Symposium 210}, Poster A20, \url{http://www.stsci.edu/hst/observatory/crds/castelli_kurucz_atlas.html}
\bibitem[Dalcanton et al.(2009)]{dalcanton09} {Dalcanton}, J.~J., {Williams}. B.~F., {Seth}, A.~C., et al. 2009, {\apjs}, 183, 67
\bibitem[de Vaucouleurs et al.(1991)]{devaucouleurs91} {de Vaucouleurs}, G., {de Vaucouleurs}, A., {Corwin}, H.~G., Jr., et al. 1991, Third Reference Catalogue of Bright Galaxies, Version 3.9 (New York: Springer)
\bibitem[Dong et al.(2015)]{dong15} {Dong}, S., {Katz}, B., {Kushnir}, D., \& {Prieto}, J.~L. 2015, {\mnras}, 454, L61
\bibitem[Diamond et al.(2015)]{diamond15} {Diamond}, T.~R., {Hoeflich}, P., \& {Gerardy}, C.~L. 2015, {\apj}, 806, 107
\bibitem[Fesen et al.(2007)]{fesen07} {Fesen}, R.~A., {Hoeflich}, P.~A., {Hamilton}, A.~J.~S., et al. 2007, {\apj}, 658, 396
\bibitem[Fesen et al.(2015)]{fesen15} {Fesen}, R.~A., {Hoeflich}, P.~A., \& {Hamilton}, A.~J.~S. 2015, {\apj}, 804, 140
\bibitem[Fesen et al.(2017)]{fesen17} {Fesen}, R.~A., {Weil}, K.~E., {Hamilton}, A.~J., \& {Hoeflich}, P.~A. 2017, {\apj}, 848, 130
\bibitem[Fink et al.(2014)]{fink14} {Fink}, M., {Kromer}, M., {Seitenzahl}, I.~R., et al. 2014, {\mnras}, 438, 1762
\bibitem[Foley et al.(2014)]{foley14b} {Foley}, R.~J., {Fox}, O.~D., {McCully}, C., et al. 2014, {\mnras}, 443, 2887
\bibitem[Fossey et al.(2014)]{fossey14} {Fossey}, S.~J., {Cooke}, B., {Pollack}, G., {Wilde}, M., \& {Wright}, T. 2014, {CBET}, 3792, 1
\bibitem[Galbany et al.(2016)]{galbany16} {Galbany}, L., {Moreno-Raya}, M.~E., {Ruiz-Lapuente}, P., et al. 2016, {\mnras}, 457, 525
\bibitem[Gall et al.(2018)]{gall17} {Gall}, C., {Stritzinger}, M.~D., {Ashall}, C., et al. 2018, {\aap}, 611, A58
\bibitem[Gamezo et al.(2003)]{gamezo03} {Gamezo}, V.~N., {Khokhlov}, A.~M., {Oran}, E.~S., et al. 2003, {Science}, 299, 77
\bibitem[Gamezo et al.(2005)]{gamezo05} {Gamezo}, V.~N., {Khokhlov}, A.~M., \& {Oran}, E.~S. 2005, 623, 337
\bibitem[Goldhaber \& Perlmutter(1998)]{goldhaber98} {Goldhaber}, G., \& {Perlmutter}, S. 1998, {\physrep}, 207, 325
\bibitem[Goobar et al.(2014)]{goobar14} {Goobar}, A., {Johansson}, J., {Amanullah}, R., et al. 2014, {\apj}, 784, L12
\bibitem[Goobar et al.(2015)]{goobar15} {Goobar}, A., {Kromer}, M., {Siverd}, R., et al. 2015, {\apj}, 799, 106
\bibitem[Greco et al.(2012)]{greco12} {Greco}, J.~P., {Martini}, P., \& {Thompson}, T.~A. 2012, {\apj}, 757, 24
\bibitem[Hamuy \& Phillips(1995)]{hamuy95} {Hamuy}, M., \& {Phillips}, M.~M. 1995, {\aj}, 109, 1
\bibitem[Hamuy et al.(1996)]{hamuy96a} {Hamuy}, M., {Phillips}, M.~M., {Schommer}, R.~A., \& {Suntzeff}, N.~B. 1996, {\aj}, 112, 2391
\bibitem[Hengeler(2014)]{hengeler14} {Hengeler}, E.-J. 2014, Bachelor's Thesis, University of Konstanz 
\bibitem[Hoeflich(2006)]{hoeflich06a} {Hoeflich}, P. 2006, {\nphysa}, 777, 579
\bibitem[Hoeflich et al.(1998b)]{hoeflich98b} {Hoeflich}, P., {Dominguez}, I., {Thielemann}, F.~K., \& {Wheeler}, J.~C. 1998, in ESA Special Publication: LIA Colloquium 34, 429, 243
\bibitem[Hoeflich et al.(2013)]{hoeflich13} {Hoeflich}, P., {Dragulin}, P., {Mitchell}, J., et al. 2013, {Fr.Phy.}, 8, 144
\bibitem[Hoeflich et al.(2002)]{hoeflich02} {Hoeflich}, P., {Gerardy}, C.~L., {Fesen}, R.~A., \& {Sakai}, S. 2002, {\apj}, 568, 791
\bibitem[Hoeflich et al.(2006)]{hoeflich06b} {Hoeflich}, P., {Gerardy}, C.~L., {Marion}, H., \& {Quimby}, R. 2006, {NewAR}, 50, 470
\bibitem[Hoeflich et al.(2004)]{hoeflich04b} {Hoeflich}, P., {Gerardy}, C.~L., {Nomoto}, K., et al. 2004, {\apj}, 617, 1258
\bibitem[Hoeflich et al.(2017)]{hoeflich17} {Hoeflich}, P., {Hsiao}, E.~Y., {Ashall}, C., et al. 2017, {\apj}, 846, 58
\bibitem[Hoeflich \& Khokhlov(1996)]{hoeflich96} {Hoeflich}, P. \& {Khokhlov}, A. 1996, {\apj}, 457, 500
\bibitem[Hoeflich et al.(2010)]{hoeflich10} {Hoeflich}, P., {Krisciunas}, K., {Khokhlov}, A.~M., et al. 2010, {\apj}, 710, 444
\bibitem[Hoeflich \& Stein(2002)]{hoeflich02c} {Hoeflich}, P. \& {Stein}, J. 2002, {\apj}, 568, 779
\bibitem[Hoeflich et al.(1998a)]{hoeflich98a} {Hoeflich}, P., {Wheeler}, J.~C., \& {Thielemann}, F.~K. 1998, {\apj}, 495, 617
\bibitem[Hoyle \& Fowler(1960)]{hoyle60} {Hoyle}, F., \& {Fowler}, W.~A. 1960, {\apj}, 132, 565
\bibitem[Hristov et al.(2018)]{hristov18} {Hristov}, B., {Collins}, D.~C., {Hoeflich}, P., {Weatherford}, C.~A., \& {Diamond}, T.~R. 2018, {\apj}, 858, 13
\bibitem[Iben \& Tutukov(1984)]{iben84} {Iben}, I.~J., \& {Tutukov}, A.~V. 1984, {\apjs}, 54, 335
\bibitem[Jack et al.(2015)]{jack15} {Jack}, D., {Mittag}, M., {Schr\"{o}der}, K.-P., et al. 2015, {\mnras}, 451, 4104
\bibitem[Khokhlov(1995)]{khokhlov95} {Khokhlov}, A.~M. 1995, {\apj}, 449, 695
\bibitem[Kushnir et al.(2013)]{kushnir13} {Kushnir}, D., {Katz}, B., {Livne}, E., \& {Fern\'{a}ndez}, R. 2013, {\apjl}, 778, L37
\bibitem[Langanke \& Mart\'{i}nez-Pinedo(2000)]{langanke00} {Langanke}, K. \& {Mart\'{i}nez-Pinedo}, G. 2000, {\nphysa}, 673, 481
\bibitem[Leonard(2007)]{leonard07} {Leonard}, D.~C. 2007, {\apj}, 670, 1275
\bibitem[Lisewski et al.(2000)]{lisewski00} {Lisewski}, A.~M., {Hillebrandt}, W., \& {Woosley}, S.~E. 2000, {\apj}, 538, 831
\bibitem[Liu et al.(2012)]{liu12} {Liu}, Z.~W., {Pakmor}, R., {Roepke}, F.~K., et al. 2012, {\aap}, 548, A2
\bibitem[Liu et al.(2016)]{liu16} {Liu}, D.-D., {Wang}, B., {Podsiadlowski}, Ph., \& {Han}, Z.  2016, {\mnras}, 461, 3653
\bibitem[Lundqvist et al.(2013)]{lundqvist13} {Lundqvist}, P., {Mattila}, S., {Sollerman}, J., et al. 2013, {\mnras}, 435, 329
\bibitem[Lundqvist et al.(2015)]{lundqvist15} {Lundqvist}, P., {Nyholm}, A., {Taddia}, F., et al. 2015, {\aap}, 577, A39
\bibitem[Maeda(2012)]{maeda12} {Maeda}, K. 2012, {\memsai}, 83, 82
\bibitem[Maeda et al.(2010)]{maeda10a} {Maeda}, K., {Benetti}, S., {Stritzinger}, M., et al. 2010, {\nat}, 466, 82
\bibitem[Maeda et al.(2014)]{maeda14} {Maeda}, K., {Kutsuna}, M., \& {Shigeyama}, T. 2014, {\apj}, 794, 37
\bibitem[Maeda et al.(2011)]{maeda11} {Maeda}, K., {Leloudas}, G., {Taubenberger}, S., et al. 2011, {\mnras}, 413, 3075
\bibitem[Maeda et al.(2010)]{maeda10b} {Maeda}, K., {Taubenberger}, S., {Sollerman}, J., et al. 2010, {\apj}, 708, 1703
\bibitem[Maeda \& Terada(2016)]{maeda16} {Maeda}, K., \& {Terada}, Y. 2016, {IJMPD}, 25, 1630024
\bibitem[Maguire et al.(2018)]{maguire18} {Maguire}, K., {Sim}, S.~A., {Shingles}, L., et al. 2018, {\mnras}, {\it in press}
\bibitem[Maguire et al.(2013)]{maguire13} {Maguire}, K., {Sullivan}, M., {Patat}, F., et al. 2013, {\mnras}, 436, 222
\bibitem[Maguire et al.(2016)]{maguire16} {Maguire}, K., {Taubenberger}, S., {Sullivan}, M., \& {Mazzali}, P.~A. 2016, {\mnras}, 457, 3254
\bibitem[Margutti et al.(2014)]{margutti14} {Margutti}, R., {Parrent}, J., {Kamble}, A., et al. 2014, {\apj}, 790, 52
\bibitem[Marion et al.(2015)]{marion15} {Marion}, G.~H., {Sand}, D.~J., {Hsiao}, E.~Y., et al. 2015, {\apj}, 798, 39
\bibitem[Mattila et al.(2005)]{mattila05} {Mattila}, S., {Lundqvist}, P., {Sollerman}, J., et al. 2005, {\aap}, 443, 649
\bibitem[Mazzali(2000)]{mazzali00} {Mazzali}, P.~A. 2000, {\aap}, 363, 705
\bibitem[Niemeyer \& Hillebrandt(1995)]{niemeyer95} {Niemeyer}, J.~C. \& {Hillebrandt}, W. 1995, {\apj}, 452, 769
\bibitem[Nomoto(1982a)]{nomoto82a} {Nomoto}, K. 1982a, {\apj}, 253, 798
\bibitem[Nomoto(1982b)]{nomoto82b} {Nomoto}, K. 1982b, {\apj}, 257, 780
\bibitem[Nomoto et al.(1984)]{nomoto84} {Nomoto}, K., {Thielemann}, F.-K., \& {Yokoi}, K. 1984, {\apj}, 286, 644
\bibitem[Pakmor et al.(2011)]{pakmor11} {Pakmor}, R., {Hachinger}, S., {R\"{o}pke}, F.~K., \& {Hillebrandt}, W. 2011, {\aap}, 528, A117
\bibitem[Pakmor et al.(2010)]{pakmor10} {Pakmor}, R., {Kromer}, M., {R\"{o}pke}, F.~K., et al. 2010, {\nat}, 463, 61
\bibitem[Pakmor et al.(2012)]{pakmor12} {Pakmor}, R., {Kromer}, M., {Taubenberger}, S., et al. 2012, {\apjl}, 747, L10
\bibitem[Penney \& Hoeflich(2014)]{penney14} {Penney}, R. \& {Hoeflich}, P. 2014, {\apj}, 795, 84
\bibitem[P\'{e}rez-Torres et al.(2014)]{pereztorres14} {P\'{e}rez-Torres}, M.~A., {Lundqvist}, P., {Beswick}, R.~J., et al. 2014, {\apj}, 792, 38
\bibitem[Phillips(1993)]{phillips93} {Phillips}, M.~M. 1993, {\apjl}, 413, L105
\bibitem[Phillips et al.(1999)]{phillips99} {Phillips}, M.~M., {Lira}, P., {Suntzeff}, N.~B., et al. 1999, {\aj}, 118, 1766
\bibitem[Raskin et al.(2009)]{raskin09} {Raskin}, C., {Timmes}, F.~X., {Scannapieco}, E., {Diehl}, S., \& {Fryer}, C. 2009, {\mnras}, 399, L156
\bibitem[Reinecke et al.(1999)]{reinecke99} {Reinecke}, M., {Hillebrandt}, W., \& {Niemeyer}, J.~C. 1999, {\aap}, 347, 739
\bibitem[Riess et al.(1995)]{riess95} {Riess}, A.~G., {Press}, W.~H., \& {Kirshner}, R.~P. 1995, {\apj}, 438, L17
\bibitem[R\"{o}pke et al.(2006)]{roepke06} {R\"{o}pke}, F.~K., {Gieseler}, M., {Reinecke}, M., {Travaglio}, C., \& {Hillebrandt}, W. 2006, {\aap}, 453, 203
\bibitem[Sand et al.(2016)]{sand16} {Sand}, D.~J., {Hsiao}, E.~Y., {Banerjee}, D.~P.~K., et al. 2016, {\apjl}, 822, L16
\bibitem[Sato et al.(2015)]{sato15} {Sato}, Y., {Nakasato}, N., {Tanikawa}, A., et al. 2015, {\apj}, 807, 105
\bibitem[Shappee et al.(2013)]{shappee13} {Shappee}, B.~J., {Stanek}, K.~Z., {Pogge}, R.~W., \& {Garnavich}, P.~M. 2013, {\apjl}, 762, L5
\bibitem[Shen(2015)]{shen15} {Shen}, K.~J. 2015, {\apjl}, 805, L6
\bibitem[Shen et al.(2012)]{shen12} {Shen}, K.~J., {Bildsten}, L., {Kasen}, D., \& {Quataert}, E. 2012, {\apj}, 748, 35
\bibitem[Siverd et al.(2015)]{siverd15} {Siverd}, R.~J., {Goobar}, A., {Stassun}, K.~G., \& {Pepper}, J. 2015, {\apj}, 799, 105
\bibitem[Spyromilio et al.(2004)]{spyromilio04} {Spyromilio}, J., {Gilmozzi}, R., {Sollerman}, J., et al. 2004, {\aap}, 426, 547
\bibitem[Srivastav et al.(2016)]{srivastav16} {Srivastav}, S., {Ninan}, J.~P., {Kumar}, B., et al. 2016, {\mnras}, 457, 1000
\bibitem[Stritzinger et al.(2015)]{stritzinger15} {Stritzinger}, M.~D., {Valenti}, S., {Hoeflich}, P., et al. 2015, {\aap}, 573, A2
\bibitem[Tanikawa et al.(2015)]{tanikawa15} {Tanikawa}, A., {Nakasato}, N., {Sato}, Y., et al. 2015, {\apj}, 807, 40
\bibitem[Telesco et al.(2015)]{telesco15} {Telesco}, C.~M., {Hoeflich}, P., {Li}, D., et al. 2015, {\apj}, 798, 93
\bibitem[Vallely et al.(2016)]{vallely16} {Vallely}, P., {Moreno-Raya}, M.~E., {Baron}, E., et al. 2016, {\mnras}, 460, 1614
\bibitem[Wang \& Han(2012)]{wang12} {Wang}, B., \& {Han}, Z. 2012, {\nar}, 56, 122
\bibitem[Webbink(1984)]{webbink84} {Webbink}, R.~F. 1984, {\apj}, 277, 355
\bibitem[Welty et al.(2014)]{welty14} {Welty}, D.~E., {Ritchey}, A.~M., {Dahlstrom}, J.~A., \& {York}, D.~G. 2014, {\apj}, 792, 106
\bibitem[Whelan \& Iben(1973)]{whelan73} {Whelan}, J., \& Iben, I.~J. 1973, {\apj}, 186, 1007
\bibitem[Wheeler et al.(1975)]{wheeler75} {Wheeler}, J.~C., {Lecar}, M., \& {McKee}, C.~F. 1975, {\apj}, 200, 145
\bibitem[Zeldovich(1970)]{zeldovich70} {Zeldovich}, Ya.~B. 1970, {\aap}, 5, 84
\bibitem[Zheng et al.(2014)]{zheng14} {Zheng}, W., Shivvers, I., Filippenko, A.~V., et al. 2014, {\apjl}, 783, L24
\bibitem[Zingale et al.(2005)]{zingale05} {Zingale}, M., {Woosley}, S.~E., {Rendleman}, C.~A., {Day}, M.~S., \& {Bell}, J.~B. 2005, {\apj}, 632, 1021
\end{thebibliography}
\end{document}